\documentclass[a4paper,fleqn,usenatbib]{mnras}

\usepackage{mathptmx}

\usepackage[T1]{fontenc}
\usepackage{ae,aecompl}

\usepackage{graphicx}	
\usepackage{amsmath}	
\usepackage{amssymb}	
\usepackage{wasysym}           
\usepackage{epstopdf}
\usepackage{mathrsfs}
\usepackage{natbib}
\usepackage{color}
\usepackage{hyperref}

\title[Mass Ejection in Failed Supernovae]{A Physical Model of Mass Ejection in Failed Supernovae}

\author[Coughlin et al.]{
Eric R. Coughlin,$^{1}$\thanks{email: eric\_coughlin@berkeley.edu}\thanks{Einstein fellow}
Eliot Quataert,$^{1}$
Rodrigo Fern\'andez$^{2}$,
Daniel Kasen$^{1,3}$
\\
$^{1}$Astronomy Department and Theoretical Astrophysics Center, University of California, Berkeley, CA 94720, USA \\
$^{2}$ Department of Physics, University of Alberta, Edmonton, AB T6G 2E1, Canada \\
$^3$ Nuclear Science Division, Lawrence Berkeley National Laboratory, Berkeley, CA 94720
}

\date{Accepted XXX. Received YYY; in original form ZZZ}

\pubyear{2017}

\begin{document}
\label{firstpage}
\pagerange{\pageref{firstpage}--\pageref{lastpage}}
\maketitle

\begin{abstract}
During the core collapse of massive stars, the formation of the protoneutron star is accompanied by the emission of a significant amount of mass-energy ($\sim 0.3 \, M_{\odot}$) in the form of neutrinos. 
This mass-energy loss generates an outward-propagating pressure wave that steepens into a shock near the stellar surface, potentially powering a weak transient associated with an otherwise-failed supernova.
We analytically investigate this mass-loss-induced wave generation and propagation.  Heuristic arguments provide an accurate estimate of the amount of energy contained in the outgoing sound pulse. We then develop a general formalism for analyzing the response of the star to centrally concentrated mass loss in linear perturbation theory. 
To build intuition, we apply this formalism to polytropic stellar models, finding qualitative and quantitative agreement with simulations and heuristic arguments.  We also apply our results to realistic pre-collapse massive star progenitors (both giants and compact stars).   Our analytic results for the sound pulse energy, excitation radius, and steepening in the stellar envelope are in good agreement with full time-dependent hydrodynamic simulations.  We show that {prior} to the sound pulses arrival at the stellar photosphere, the photosphere has already reached velocities $\sim 20-100 \%$ of  the local sound speed, thus likely modestly decreasing the stellar effective temperature prior to the star disappearing.   Our results provide important constraints on the physical properties and observational appearance of failed supernovae.
\end{abstract}




\section{Introduction}
Many massive stars ($\gtrsim 8M_{\odot}$) end their lives in fantastic explosions.  However, successfully simulating core-collapse supernovae -- and discerning the underlying mechanism for the explosion itself -- has proven to be extremely difficult. Indeed, the simplest, original notion that a shock generated during the formation of the neutron star could propagate through and unbind the stellar envelope was shown to fail in most cases \citep{bethe90}, and even the ``revival'' of such a shock through neutrino heating is likely insufficient for the most energetic supernovae \citep{janka16}. 

While the consensus of what leads to a successful supernova is still far from established, it is clear that the properties of the progenitor 
play an important role in determining the likelihood of a successful explosion.  In particular, stars with a more compact iron core and a shallower density profile outside the iron core are more difficult to explode (e.g., \citealt{oconnor11,Ertl2016}).   It therefore seems likely that, while many stars do succeed in expelling their outer layers in a supernova, some may actually be incapable of doing so. These ill-fated stars would reach the onset of core-collapse, form a neutron star, and  collapse to a stellar-mass black hole through the continued accretion of the surrounding star. 

As pointed out by \citet{kochanek08}, it might be possible to detect such disappearing stars as just that: a point on the sky where a star used to be (and no intervening supernova).   It is also plausible, however, that failed supernovae manifest themselves observationally as more than just a disappearing star.  
In this paper we analyze a mechanism by which nominally failed supernovae can in fact eject some mass, potentially powering a weak transient.   This mechanism was first suggested by \citet{nadezhin80} and investigated further by \citet{lovegrove13} and \citet{lovegrove17}: 
the proto-neutron star phase of core-collapse
generates a prodigious, short-lived, mass-energy loss in the form of neutrinos. These neutrinos stream out of the star with essentially no direct interaction with the outer envelope on a timescale of a few seconds, and carry with them $\sim few \times 0.1M_{\odot}$ of mass \citep{burrows88}. This mass loss creates a nearly-instantaneous drop in the gravitational field according to the outer layers of the star, where the free-fall time is much greater than the few seconds over which the neutrinos are radiated. This causes the now-overpressured star to respond dynamically and expand, generating an outwardly propagating sound pulse that steepens into a shock in the outer layers of the star.

\citet{nadezhin80} performed the first hydrodynamical simulations of this mechanism.
\citet{lovegrove13} and \citet{lovegrove17} reinvestigated the hydrodynamics of this process and assessed -- predominantly numerically and only for red super-giant progenitors -- the energetics and appearance of this relatively low-energy expulsion of the stellar envelope.  \citet{fernandez17} (hereafter F17) performed a numerical study of mass ejection for a wider range of stellar progenitors, including red supergiants, blue supergiants and Wolf-Rayet stars. Interestingly, one case of a failed supernovae associated with a red super-giant progenitor may have already been found \citep{adams17b,adams17}. 

Despite the fact that the first simulations of this effect date back nearly 40 years, there is still no analytic understanding of the generation and evolution of the shock that ultimately produces the ejection of some of the stellar envelope in failed SNe. 
Our primary goal here is to provide such an understanding, which in turn will provide useful constraints on the ability of mass loss during failed SNe to generate observationally detectable  transients. In Section \ref{sec:scalings} we provide the basic, phenomenological picture of the effect first identified by Nadyozhin, and we use this basic understanding to approximate the energy released in the explosion. In Section \ref{sec:general} we pursue a rigorous analysis of the effects of the mass loss on the stellar structure by performing a linear perturbation analysis; we present fundamental equations for, e.g., the velocity induced throughout the envelope in terms of the Eigenmodes of the stellar progenitor, and we derive a general expression for the energy imparted by the mass loss that agrees with the results of Section \ref{sec:scalings}. Section \ref{sec:polytrope} applies the results of Section \ref{sec:general} to polytropic stellar models, and Section \ref{sec:stars} applies our findings to more realistic stellar progenitors. We summarize our results and conclude in Section \ref{sec:conclusions}.   Appendix \ref{sec:appendix} compares our linear perturbation theory predictions for the mass-loss-induced evolution of a polytrope to the results of a 1-D, Lagrangian hydrodynamics code. The early evolution is identical between the two approaches, validating the analytic methods employed throughout the bulk of this paper.

\section{Physical Picture and Estimates}
\label{sec:scalings}
At the onset of core collapse, the formation of the proto-neutron star is accompanied by the loss of $\sim 0.1-0.5 M_{\odot}$ of mass from the central core of the star. To first approximation, this mass loss results in an outward motion of the stellar envelope owing to the reduced gravitational field, and the velocity profile generated is approximately given by the solution to the radial momentum equation. In the limit that the mass loss occurs impulsively, the resultant velocity profile is simply\footnote{There are relativistic corrections to this expression that arise from the fact that the change in the gravitational field is conveyed to the outer layers of the star at the speed of light; however, these corrections are always small (or order $v_{ff} / c$, $v_{ff}$ being the free-fall velocity), and we will proceed with our Newtonian approximations.}

\begin{equation}
v = \frac{G\delta{M}}{r^2}t, \label{vinit}
\end{equation}
where $\delta{M}$ is the mass lost to neutrinos, $t$ is time after the core-collapse, and $r$ is radial distance from the center of the star. We have assumed, as we will throughout the remainder of this paper, that the star is spherically symmetric.

Equation \eqref{vinit} gives the velocity profile that develops  throughout the envelope following the mass loss. This applies in the regions for which the mass loss is effectively instantaneous, i.e., for which the local dynamical time is longer than the time over which the neutrino binding energy is radiated.   However, the central regions of the star instead collapse onto the proto-neutron star, which clearly violates the scaling given by Equation \eqref{vinit}. Thus, while Equation \eqref{vinit} gives the initial, dynamical response of the envelope, there is an additional, pressure-mediated reaction that conveys to the outer regions of the envelope that the central regions are infalling. This additional response is in the form of a pressure wave that travels outward into the envelope at the local sound speed; this pressure wave ``tells'' the outward-moving material to stop expanding.

We can use this physical understanding to predict the energy contained in the sound wave as it travels out into the envelope: since the sound pulse propagates at the local sound speed, the velocity everywhere in the star reaches a radially-dependent, maximum value of 

\begin{equation}
v_{max} = \frac{G\delta{M}\tau_{sc}(r)}{r^2}, \label{vmax}
\end{equation}
where 

\begin{equation}
\tau_{sc}(r) = \int_{r_c}^r\frac{d\tilde{r}}{c_s(\tilde{r})}
\label{eq:tsc}
\end{equation}
is the sound crossing time from the inner radius $r_c$ out to radius $r$, and $c_s(r)$ is the local sound speed. After the time $\tau_{sc}$, the outward moving mass shells move back inwards toward the forming neutron star, and thus the change in velocity induced by the sound pulse is approximately given by Equation \eqref{vmax}. Therefore, as the sound pulse moves outward into the envelope, its energy grows as

\begin{equation}
\Delta{E} \sim 0.5 \int_{r_c}^{r} \frac{G^2\delta{M}^2\tau_{sc}(r)^2}{r^4}d M_r \sim \frac{G\delta M^2}{2 \alpha} \label{Eapp},
\end{equation}
where 
$\alpha$ is the pressure scale height for radii $\sim {\rm few} \, r_c$ ($\tau_{sc}$ is defined relative to $r_c$ [eq. \ref{eq:tsc}] so the energy input is dominated by radii a few times larger; see Section \ref{sec:stars}). 
For the proto-neutron star problem of interest, the radius $r_c$ is roughly set by smaller of the radius with a free-fall time of a few seconds (the neutrino diffusion time of a proto-neutron star) or the radius enclosing $\sim 2-3 \, M_\odot$ in the progenitor, which is the region that collapses to form a black hole.  This is $r_c \sim 1.5 \times 10^9$ cm for typical progenitors (see 
F17 and Section \ref{sec:stars}).  Setting $\delta{M} = 0.2 M_{\odot}$, $\alpha = 3 \times r_c \simeq 4.5 \times 10^9$ cm, we find
\begin{equation}
\Delta E \sim  10^{48} \text{ erg}.
\end{equation}
The more detailed calculations for realistic stellar progenitors in Section \ref{sec:stars} corroborate this estimate. 

The energy $\Delta E$ estimated in equation \ref{Eapp} is contained in the sound pulse as it propagates through the stellar envelope. If the sound pulse remained everywhere linear, the sound wave containing this energy would reflect off of the stellar surface, resulting in an overall increase in the energy of the star.

However, one can show from linear theory that the power carried by the sound pulse is approximately conserved (\citealt{dewar70,ro17}), such that the velocity immediately behind the sound pulse satisfies

\begin{equation}
4\pi\rho v^2 r^2 c_s \simeq const \sim E_c\sqrt{4\pi G \rho(r_c)}, \label{vapp}
\end{equation}
where $E_c = G \delta M^2/\alpha$, $\rho(r_c)$ is the mean density interior to $r_c$ and the right-hand side of this expression results from equation \ref{Eapp} and applying the left-hand side near the radius $r_c$. Equation \ref{vapp} implies that near the surface of the star where the density and the sound speed become small (or, in the case of a red supergiant, near the boundary of the hydrogen envelope where the density drops precipitously), the velocity will increase to the point where it becomes supersonic, generating a shock. In particular, using the above expression for the velocity, dividing by the sound speed, and performing some simple algebraic manipulations shows that the Mach number is given by

\begin{equation}
\mathscr{M} \sim \frac{\delta{M}}{M_{in}}\sqrt{\frac{1}{y x^2 c^3}}, \label{mach}
\end{equation}
where $M_{in}$ is the mass enclosed within $r_c$, and $y$, $x$, and $c$ are the density, radius, and sound speed normalized by their respective values at $\sim r_c$.

From Equation \eqref{mach} it is evident that the flow will be subsonic throughout the majority of the envelope.   In the central regions of the star where the pressure and density are roughly constant, the Mach number can actually decrease $\propto 1/r$ due to the geometrical dilution of the energy contained in the outgoing sound wave.

It is only once the pressure disturbance starts to reach the stellar surface, where the sound speed and density decline appreciably, that the Mach number will start to increase. The location where the Mach number equals unity -- and therefore results in shocks -- cannot be written down explicitly, but for a given stellar model and $\delta M$, this equation can be solved easily for the approximate location at which a shock will form. We will investigate this more in subsequent sections.

Finally, we note that the Mach number given by Equation \eqref{mach} describes the fluid velocity immediately behind the outgoing pressure wave. However, from Equation \eqref{vinit} we see that there is a radial dependence of the initial acceleration within the star, meaning that mass shells at smaller radii catch up to those at larger radii. It is therefore possible that this alone could cause shell crossings (i.e., shocks) during the initial, dynamical evolution of the envelope, prior to the sound pulse reaching large radii.  This is the most likely to occur near the stellar surface where the sound speed is small.   We will investigate this in more detail in Sections \ref{sec:polytrope} and \ref{sec:stars}.

\section{General Solutions in the perturbative limit}
\label{sec:general}
The above analysis provides a rough understanding of the physical mechanism and energetics associated with mass ejection by neutrino radiation in otherwise failed SNe.

While this picture is qualitatively and, to a lesser extent, quantitatively consistent with simulations, a more accurate analysis is possible because the change in mass is small compared to the total stellar mass itself. The initial response of the stellar envelope can thus be accurately calculated using linear perturbation theory.  The linear approximation breaks down when the sound pulse steepens into a shock in the outer, low density parts of the star via Equation \eqref{mach}.  However, the energy in the pulse is largely determined in the linear phase that we now proceed to calculate.   Moreover, we can use the linear results to accurately determine where in the stellar envelope the sound pulse transitions into a shock.

\subsection{Equations}
The evolution of the stellar envelope is described by the continuity equation, the radial momentum equation, and the gas energy equation, which respectively read 

\begin{equation}
\frac{\partial m}{\partial t}+v\frac{\partial m}{\partial r} = 0,
\end{equation}
\begin{equation}
\frac{\partial v}{\partial t}+v\frac{\partial v}{\partial r}+\frac{1}{\rho}\frac{\partial p}{\partial r} = -\frac{G M}{r^2},
\end{equation}
\begin{equation}
\frac{\partial K}{\partial t}+v\frac{\partial K}{\partial r} = 0,
\end{equation}
where $m$ is the stellar mass contained within radius $r$, $v$ is the radial velocity, $p$ and $\rho$ are the respective gas pressure and density, $M$ is the total mass contained within radius $r$ (which differs from $m$ if there is a point mass, corresponding in our case to a central neutron star or black hole), and $K = p/\rho^{\gamma}$ is the specific entropy of the gas with $\gamma$ the adiabatic index. We assumed that the gas obeys an adiabatic equation of state, and for simplicity we set the adiabatic index to a constant (though the inclusion of a radially-dependent $\gamma$ -- which could certainly be relevant for some stars -- is straightforward).

At $t = 0$, the interior of the star collapses to a neutron star, which radiates a time-dependent amount of mass $M_{\nu}(t)$ in the form of neutrinos. Letting $M_{\nu}$ be small relative to the total mass of the star (which is true in all astrophysical situations), we linearize the above three equations in that small quantity and keep only first-order terms; quantities with a subscript 0 will refer to the initial state of the star prior to the mass loss, while those with a subscript 1 refer to time and space-dependent perturbations resulting from the time-dependent gravitational field. We also Laplace transform the equations in anticipation of an instantaneous and conceivably discontinuous mass loss at $t = 0$, and we denote Laplace-transformed variables by tildes, i.e., 

\begin{equation}
\tilde{v}_1(s,r)=\int_0^{\infty}v_1(t,r)e^{-st}dt.
\end{equation}
Doing so yields, self-consistently and in line with expectations, the equation of hydrostatic equilibrium for the unperturbed quantities: 

\begin{equation}
\frac{1}{\rho_0}\frac{\partial p_0}{\partial r} = -\frac{Gm_0}{r^2}.
\end{equation}
Making a number of algebraic manipulations we find the following fundamental equation for the linear mass flux $F_1 = r^2\rho_0 v_1$:

\begin{equation}
s_*^2\tilde{F}_1-\mathcal{L}[\tilde{F}_1] = -\frac{s_*}{\sqrt{4\pi G\rho_c}}\rho_0 G\tilde{M}_{\nu}, \label{fluxeq}
\end{equation} 
where

\begin{equation}
\begin{split}
\mathcal{L}[\tilde{F}_1] = & \left(\rho+\frac{\xi^2}{\gamma}\frac{\partial}{\partial \xi}\left[\frac{p}{\rho \xi^2}\frac{\partial}{\partial \xi}\ln K_0\right]\right)\tilde{F}_1
\\ & +\rho \xi^2\frac{\partial}{\partial \xi}\left[\frac{p}{\rho^2\xi^2}\frac{\partial\tilde{F}_1}{\partial \xi}\right]. \label{Loperator}
\end{split}
\end{equation}
We have also non-dimensionalized this equation by introducing the following variables:

\begin{equation}
\begin{split}
\rho(r) = & \frac{\rho_0(r)}{\rho_c},\quad p(r) = \frac{p_0(r)}{p_c}, \quad s_* = \frac{s}{\sqrt{4\pi G\rho_c}},
\\ & \xi = \frac{r}{\alpha}, \quad \alpha^2 = \frac{\gamma p_c}{4\pi G\rho_c^2},
\end{split}
\end{equation}
where $\rho_c$ and $p_c$ are the central density and pressure, respectively. Note that $\alpha$ is the pressure scale height introduced in Section \ref{sec:scalings}.  In what follows we will also use a dimensionless mass loss 
\begin{equation}
\delta m = \frac{\delta M}{4 \pi \rho_c \alpha^3}.
\end{equation}
In deriving equation \eqref{fluxeq}, we assumed that the initial mass flux in the star -- and hence the initial velocity -- was zero. While this assertion is in line with the fact that the star was initially in hydrostatic equilibrium, setting the initial velocity to zero also prevents one from self-consistently allowing the inner regions of the envelope to fall onto the protoneutron star. One can surmount this issue by assuming that there is a small, but non-zero, $v_0$ that is established during the initial loss of pressure support.  By small we mean that we continue to ignore terms such as $v_0 \partial v_1/\partial r$ in the momentum equation that would formally be present at linear order given a non-zero $v_0$; this is equivalent to assuming that the initial radial velocity is small compared to the sound speed. 

We have investigated the consequences of permitting such a non-zero $v_0$, which results in a closer alignment between the solutions we seek here and simulations. We have found that the results generally disagree by a factor of two at most for the properties of the outgoing sound pulse. Therefore, because it makes for a simpler analysis and does not significantly alter our key results, we will henceforth proceed by assuming that the initial mass flux throughout the star is zero.  

\subsection{Solutions}
We solve Equation \eqref{fluxeq} by expanding the flux in the Eigenmodes of the operator $\mathcal{L}$. In particular, we write 

\begin{equation}
\tilde{F}_1=\sum_{\sigma}c_{\sigma}A_{\sigma}(\xi),
\end{equation}
where $c_\sigma$ is a coefficient that depends on $s_*$, and $A_\sigma$ is an Eigenfunction that satisfies

\begin{equation}
\mathcal{L}[A_\sigma] = -\sigma^2A_\sigma.
\end{equation}
Because $\mathcal{L}$ is in Sturm-Liouville form, the $A_\sigma$ represent an orthogonal basis and can be normalized such that

\begin{equation}
\int \frac{A_{\sigma}A_\beta d\xi}{\rho\xi^2} = \delta_{\sigma\beta}.
\label{eq:norm}
\end{equation}
Note that because the $A_\sigma$ are eigenfunctions of the linear mass flux $\propto \rho \xi^2 v_1$ the term in the denominator in equation \ref{eq:norm} cancels and so there is no concern about divergence as either $\xi \rightarrow 0$ or $\rho \rightarrow 0$.  

In our linear theory, the Eigenmodes ensure the regularity of the solutions at the stellar surface by satisfying $A_{\sigma}(\xi_1) = 0$, where $\xi_1$ is the value of $\xi$ where $\rho(\xi_1) = 0$ (i.e., the surface). A nonlinear approach would account for the fact that the surface can expand, and thus the outer boundary condition should take place on this moving mass shell. While this latter boundary condition is more physical, our linear approach that imposes the boundary condition on the unperturbed star will still give a good approximation to the initial formation and propagation of the sound pulse through the stellar envelope. 

We also require that the Eigenmodes satisfy $A_{\sigma}(\xi_c) = 0$, where $\xi_c$ is some inner radius greater than zero. Because we are modeling the formation of the protoneutron star as a point mass, we cannot extend our solutions all the way to the origin, which would be the location of the inner boundary condition in standard, stellar pulsation theory.   Physically, the inner boundary here is assumed to be just outside the region that collapses to produce the proto-neutron star, where it is reasonable to assume a nearly hydrostatic solution during the time over which the neutrino radiation occurs.   

Inserting our series solution over the Eigenmodes and exploiting their orthogonality yields the coefficients $c_\sigma$. We adopt an exponential form for the mass loss, so that 
\begin{equation}
M_{\nu}(t) = -\delta M\left(1-e^{-\omega_* \tau}\right), \label{dmexp}
\end{equation}
where $\tau = t \sqrt{4 \pi G \rho_c}$, $\omega_* = \omega/\sqrt{4 \pi G \rho_c}$ and $\omega^{-1}$ characterizes the timescale over which the mass is lost.  With this choice of $M_\nu(t)$, we find for the mass flux

\begin{equation}
F_1 = \frac{\delta{M}\sqrt{4\pi G\rho_c}}{4\pi}\sum_{\beta}\frac{\omega_*}{\omega_*^2+\beta^2}R_{\beta}(\tau)A_{\beta}(\xi), \label{fluxex}
\end{equation}
where

\begin{equation}
R_{\beta} = \left\{e^{-\omega_*\tau}-\cos\beta\tau+\frac{\omega_*}{\beta}\sin\beta\tau\right\}\int_{\xi_c}^{\xi_1}\frac{A_{\beta}}{\xi^2}d\xi. \label{Rbetam}
\end{equation}

The temporal evolution implied by Equations \eqref{fluxex} and \eqref{Rbetam} agrees qualitatively with what we expect: the flux is zero at $\tau = 0$ when the system is in hydrostatic equilibrium. The neutrino induced mass loss then generates a time-dependent velocity that, when $\omega_* \lesssim \beta$, scales as
$R_{\beta} \propto \omega_*\tau^2$, which results from a Taylor expansion with $\omega_*\tau \ll 1$.  On the other hand, if $\omega_* \gg \beta$, then the last term in braces in Equation \eqref{Rbetam} quickly dominates over the other two, and we have $R_{\beta} \propto \tau$, again Taylor expanding for early times.  In fact, in this regime ($\omega_* \gg \beta$ and $\beta \tau \ll 1$), it is straightforward to show that equation \ref{fluxex} reduces to the much simpler equation \ref{vinit}.

\subsection{Energy}
We can manipulate the energy, momentum, and gas energy equations to yield the following conservation law for the total energy:

\begin{equation}
\frac{\partial\mathscr{E}}{\partial t}+\frac{1}{r^2}\frac{\partial}{\partial r}\left[r^2\mathscr{F}\right] = -\frac{GM_{\nu}(t)\rho_0 v}{r^2}, \label{energysph}
\end{equation}
where the energy density $\mathscr{E}$ and energy flux $\mathscr{F}$ are

\begin{multline}
\mathscr{E} = \frac{1}{2}\bigg\{\rho_0 v^2+\rho_0 c_s^2\left(\frac{\rho_1}{\rho_0}\right)^2 - \frac{Gm_1^2}{4\pi r^4}\\ 
-\frac{m_1^2}{4\pi r^2\rho_0}\frac{\partial}{\partial r}\left[\frac{p_0}{4\pi r^2\rho_0}\frac{\partial}{\partial r}\ln K_0\right]\bigg\} \label{energydensity}
\end{multline}
and

\begin{equation}
\mathscr{F} = c_s^2\rho_1 v,
\end{equation}
$c_s^2 = \gamma p_0/\rho_0$ being the square of the local sound speed. 

The first two terms in the expression for $\mathscr{E}$ are from the kinetic energy and the thermal energy, respectively. The third term is the change in the gravitational potential energy of the fluid generated by introducing a mass perturbation $m_1$. The origin of the fourth term is less obvious, but it can be interpreted as the energy liberated by the buoyant advection of matter, where the buoyancy comes from the entropy gradient, $\partial K_0/\partial r$, in the stellar interior. When the entropy gradient is positive, this term creates an additional energy sink, and effectively translates to a greater difficulty in moving mass shells upward in the atmosphere. The term on the right-hand side of Equation \eqref{energysph} is the work done by the excess pressure force produced by the change in mass due to neutrinos, and is ultimately what drives the time-dependent evolution of the stellar envelope.   Note that this does not show up in the self-gravity term $\propto m_1^2$ because in our formulation the neutrino-induced mass change applies to the central point mass, not the gas in the star (only the latter is described by $m_1$, $v_1$, etc.).   

Multiplying Equation \eqref{energysph} by $4\pi r^2$ and integrating from $r_c$ to $r_1$ yields

\begin{equation}
\frac{\partial E_{tot}}{\partial \tau}
= \delta{E}(t), \label{etot1}
\end{equation}
where

\begin{equation}
E_{tot}= \int_{r_c}^{r_1}4\pi r^2\mathscr{E}dr
\end{equation}
is the total integrated energy contained in the star (we used the fact that the flux vanishes at the inner and outer radii). Using the solution for the flux in terms of the Eigenmodes, we can show that

\begin{multline}
\delta{E}(t)=E_c\left(1-e^{-\omega_*\tau}\right)\sum_{\beta}\frac{\omega_*}{\omega_*^2+\beta^2} \\ \times\left\{e^{-\omega_*\tau}-\cos\beta\tau+\frac{\omega_*}{\beta}\sin\beta\tau\right\}\left(\int_{\xi_c}^{\xi_1}\frac{A_{\beta}d\xi}{\xi^2}\right)^2, \label{dEdtau}
\end{multline}
where

\begin{equation}
E_c = \frac{G\delta{M}^2}{\alpha}.
\end{equation}
Finally, integrating Equation \eqref{etot1} over time gives the total energy contained in the star as a function of the mass lost to neutrinos. Using the above expression for $\delta{E}(t)$, we can show that this integrated energy is

\begin{equation}
E_{tot}(\tau) = 
E_c\sum_{\beta}\left(\int_{\xi_c}^{\xi_1}\frac{A_{\beta}d\xi}{\xi^2}\right)^2\frac{C_{\beta}(\tau)}{\omega_*^2+\beta^2}, \label{Etotal}
\end{equation}
where 

\begin{multline}
C_{\beta}(\tau)=\frac{1}{2}\left(1-e^{-\omega_*\tau}\right)^2 -\frac{\omega_*}{\beta}\sin\beta\tau\left(1-e^{-\omega_*\tau}\right) \\ +\frac{\omega_*^2}{\beta^2}\left(1-\cos\beta\tau\right)
\end{multline}

Even though it is somewhat complicated in detail, Equation \eqref{Etotal} yields the general result that the total energy imparted by the mass loss is

\begin{equation}
\Delta E_{\nu} \simeq \frac{G\delta{M}^2}{\alpha},
\end{equation}
where $\alpha$ is the pressure scale height. This result confirms the order-of-magnitude estimates of Section \ref{sec:scalings}, and we see that Equation \eqref{Eapp} is accurate up to a dimensionless number related to integrals over the Eigenmodes of the star. This expression for the energy is accurate to lowest order in the velocity to the central sound speed, which is generally quite small.

Additionally, it is possible to show that the coefficients $C_\beta$ satisfy

\begin{equation}
C_{\beta}>\frac{1}{2}\left(1-e^{-\omega_*\tau}-\frac{\omega_*}{\beta}\sin\beta\tau\right)^2.
\end{equation}
The positive-definiteness of $C_{\beta}(\tau)$ illustrates that the change in energy due to the mass loss is always positive. While the increase in the energy is reasonable -- it would have been surprising if a decrease in the gravitational field of the central object resulted in a \emph{more} bound stellar envelope -- the total energy change is not time independent. We will investigate this result in more detail in Section \ref{sec:polytrope}, where we analyze these solutions in the specific case where the progenitor is a polytrope. 

\section{Polytropes}
\label{sec:polytrope}
The analysis of the preceding section was completely general and valid for any stellar progenitor. Here we focus our attention on the case when the unperturbed stellar density and pressure profiles are those of a polytrope, which allows us to explicitly solve for the Eigenmodes and, therefore, the response of the star to the neutrino mass loss. For our purposes here, a polytrope provides a simple, physically-tenable description of a stellar interior, and suffices to yield tangible results  from the analysis of Section \ref{sec:general}.   We discuss the application of these results to real stellar progenitors in \S \ref{sec:stars}.   The primary difference is that real stellar progenitors are far more stratified than a single polytrope.  Thus the polytropic model is best interpreted as an approximation to the stellar structure in the region outside the iron core that becomes the proto-neutron star but interior to any very extended hydrogen envelope. 

We will analyze the specific cases of $\gamma = 1.4$, 1.5, and 1.6 polytropes, focusing primarily on $\gamma = 1.5$.  We will also assume for simplicity that the adiabatic index of the perturbations equals the polytropic index, meaning that the entropy gradient is identically zero throughout the entirety of the star and the second term in parentheses on the right-hand side of Equation \eqref{Loperator} vanishes. 

There are three quantities that we can vary in the solutions (and would likely vary between progenitors), being the total mass lost to neutrinos, $\delta{M}$, the rate of neutrino-induced mass-loss, $\omega$, and the location of the inner radius, $r_c$. Because all quantities are linear in the mass loss, $\delta{M}$ can be scaled out of the problem, and the mass flux, velocity, density perturbation, and other first order variables can be normalized by this quantity. The other two, however, must be specified.

\begin{figure*}
\includegraphics[width=0.323\textwidth]{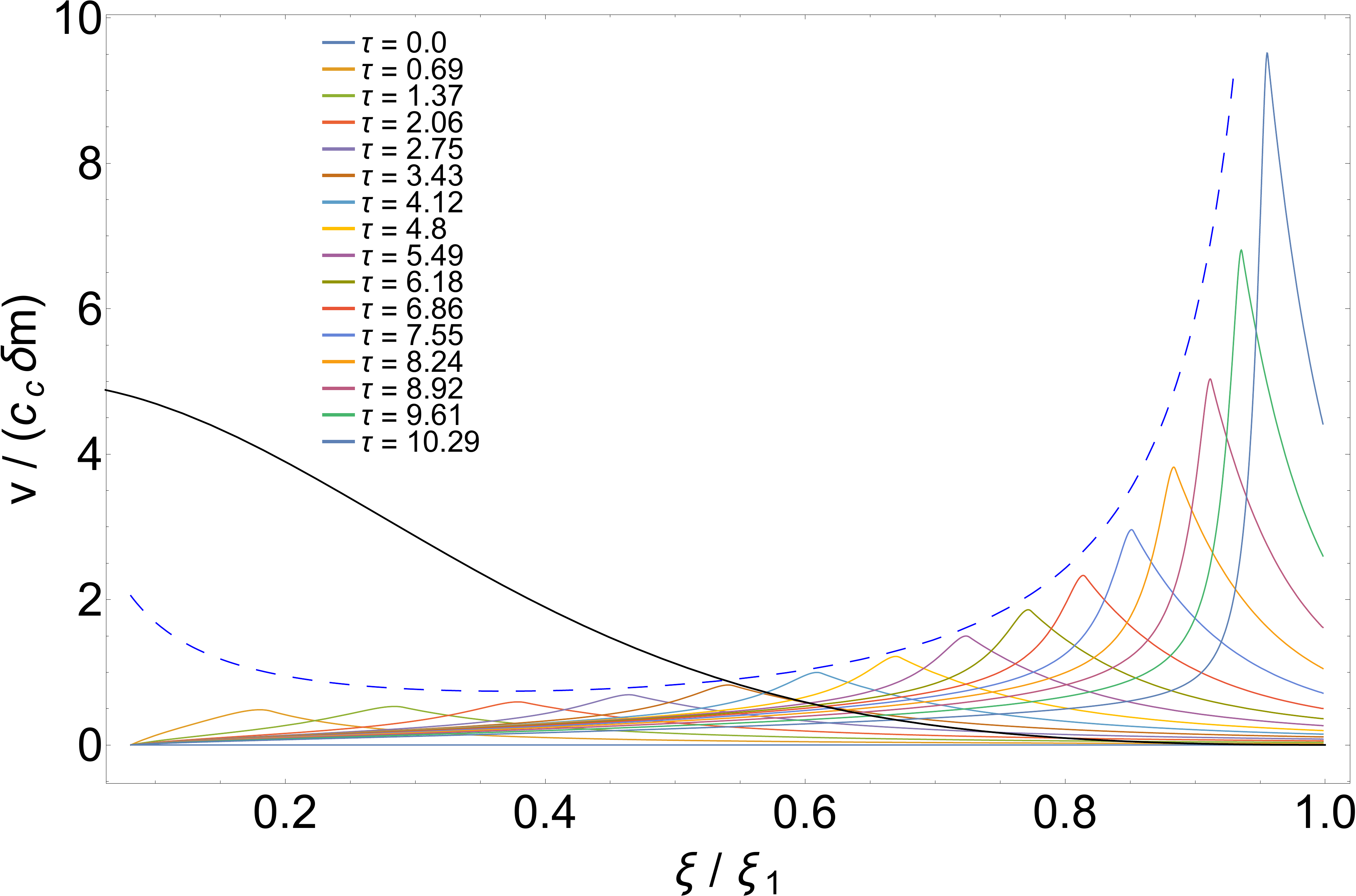}
\includegraphics[width=0.33\textwidth]{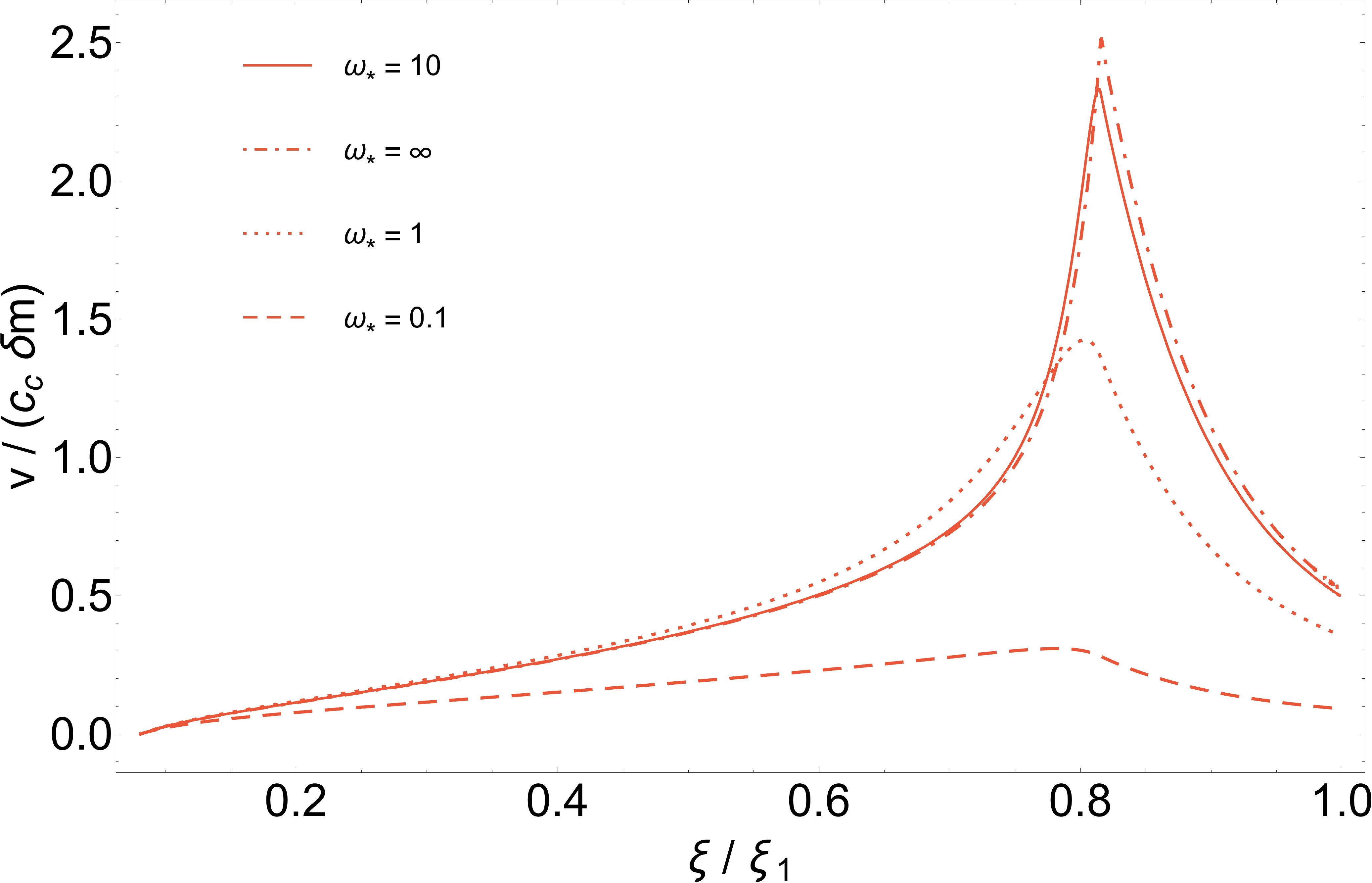}
\includegraphics[width=0.33\textwidth]{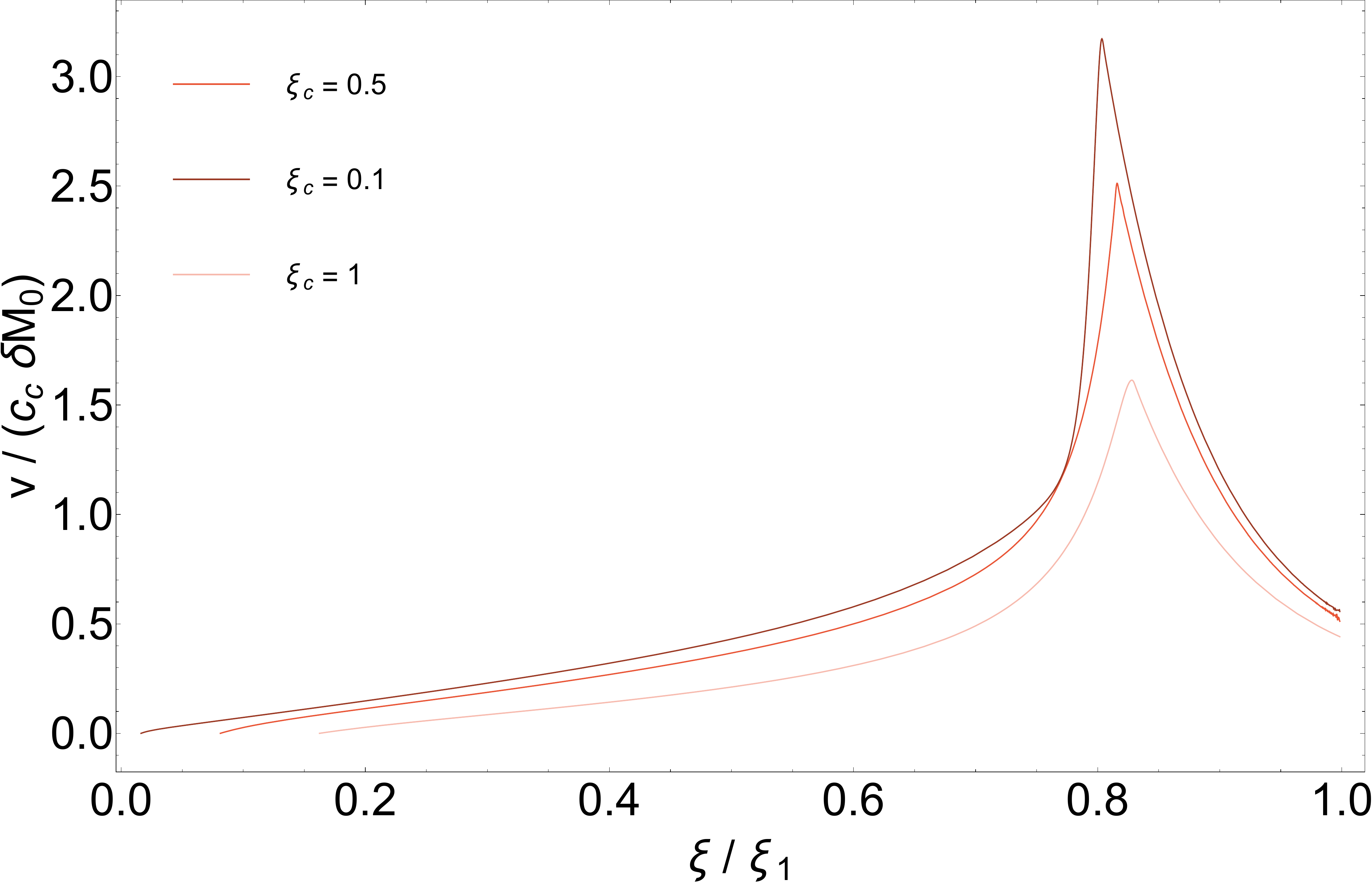}
\caption{Left: Velocity as a function of dimensionless radius $\xi / \xi_1$, where $\xi_1$ is the surface of the polytrope, in a $\gamma = 1.5$ polytrope, where different curves are at the times shown in the legend, with $\omega_* = 10$ and $\xi_c = 0.5$; the blue, dashed curve shows the estimate given by Equation \eqref{vapp}, and the thick, black curve shows the density profile of the polytrope scaled by a factor of five for clarity. Middle: The velocity profile at a time of $\tau = 6.86$ for $\omega_* = 10$ (which reproduces the red curve in the left-hand panel), $\omega_* = \infty$ (dot-dashed curve), $\omega_* = 1$ (dotted curve), and $\omega_* = 0.1$ (dashed curve). Right: The velocity profile at a time of $\tau = 6.86$ and a variable $\xi_c$, shown in the legend.}
\label{fig:velocity_radp5_omega10}
\end{figure*}

\begin{figure*}
\includegraphics[width=0.495\textwidth]{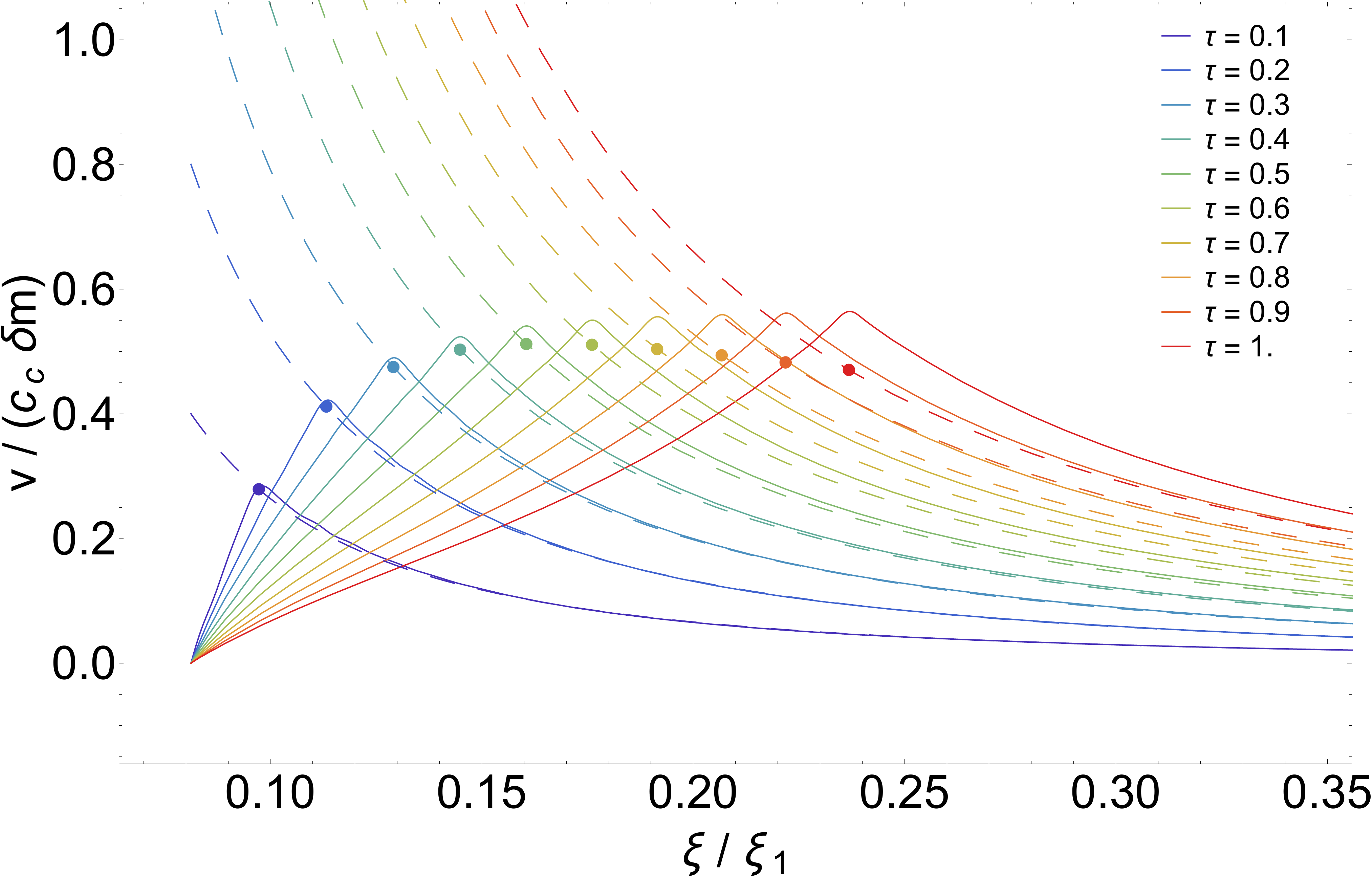}
\includegraphics[width=0.495\textwidth]{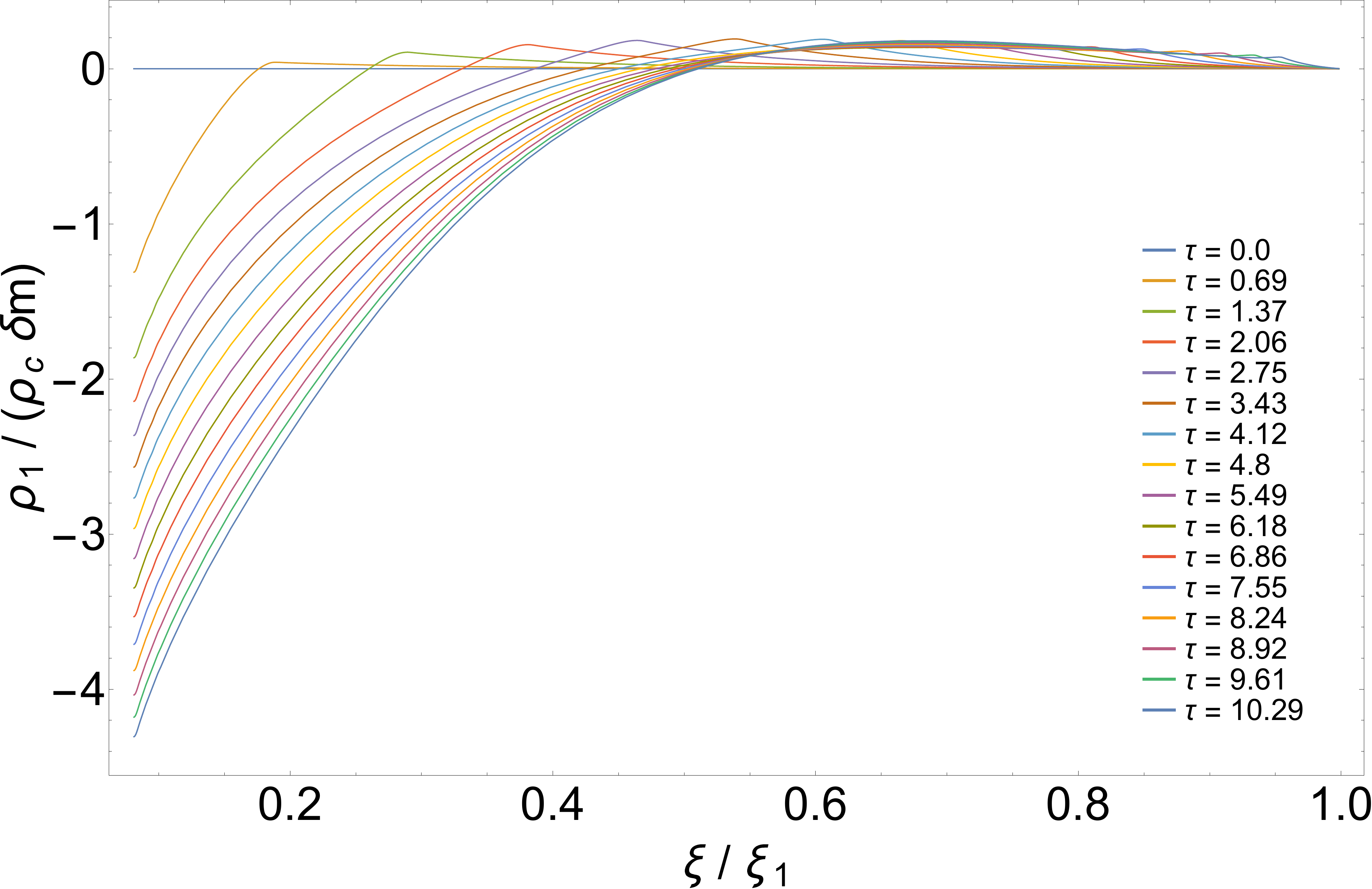}
\caption{Left: The solid lines show the velocity profile throughout a $\gamma = 1.5$ polytrope when $\omega_* = \infty$ and $\xi_c = 0.5$ at the times in the legend; the dashed lines illustrate the dynamic response $v = \tau/\xi^2$ (eq. \ref{vinit}), which holds during the early evolution and for radii greater than the radius out to which the sound pulse has propagated in time $\tau$; the points show the position of the sound pulse at time $\tau$; Right: The density profile for $\omega_* = 10$ and $\xi_c = 0.5$ throughout a $\gamma = 1.5$ polytrope at the times in the legend.  The decrease in density is because the sound pulse propagating out to large radii is a rarefaction wave.}
\label{fig:velocity_analytic}
\end{figure*}

\begin{figure*}
\includegraphics[width=0.323\textwidth]{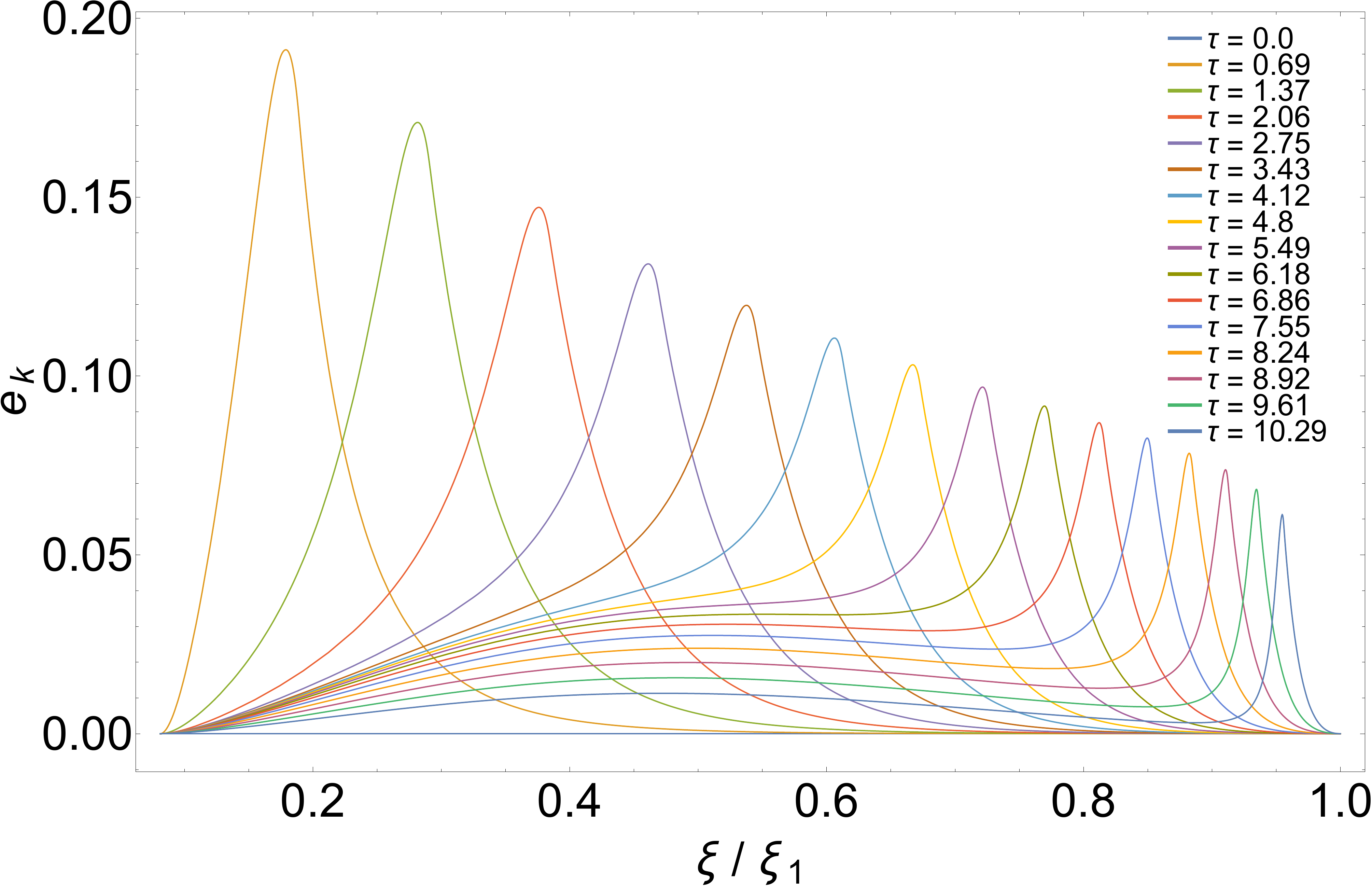}
\includegraphics[width=0.33\textwidth]{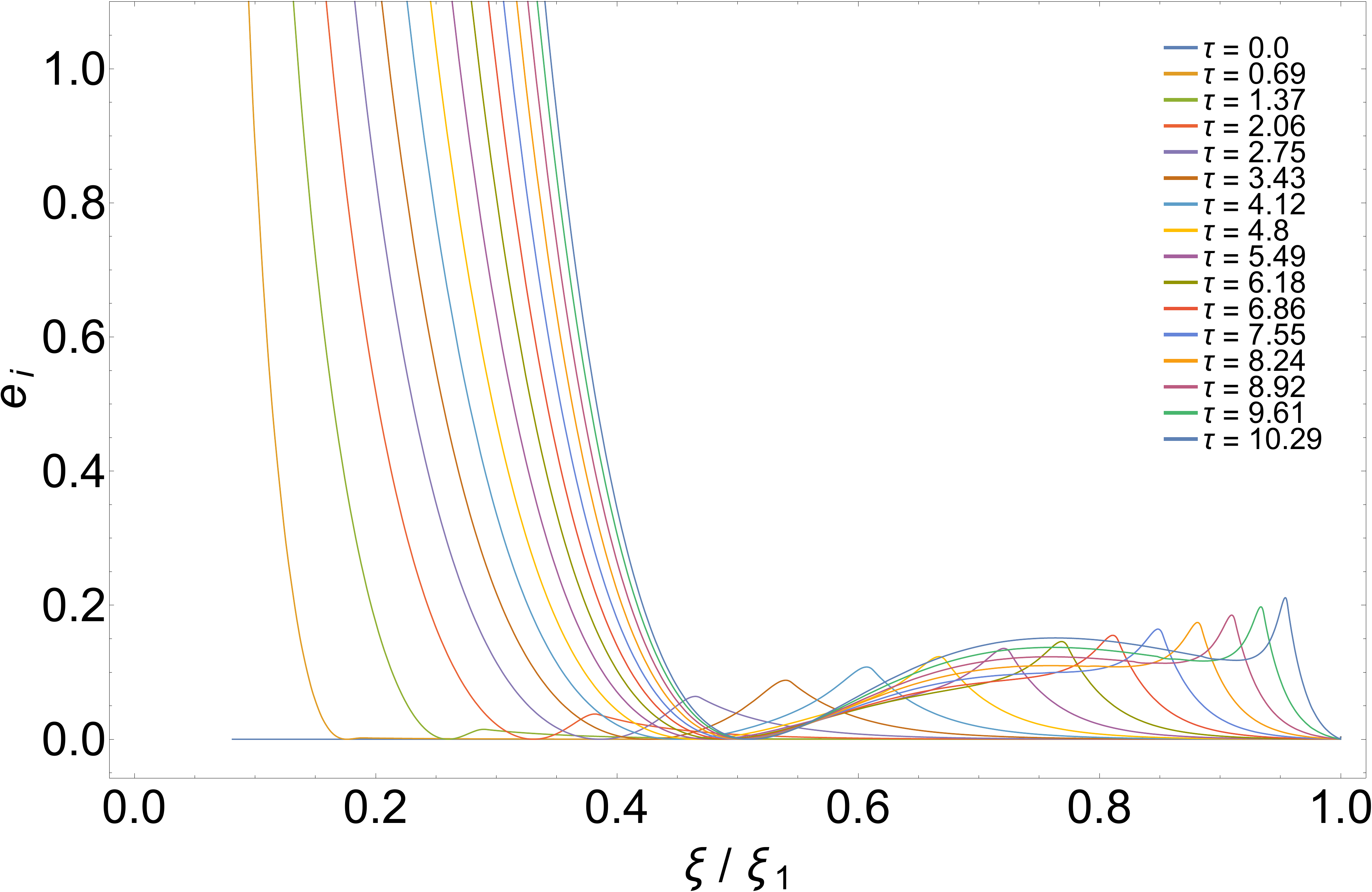}
\includegraphics[width=0.33\textwidth]{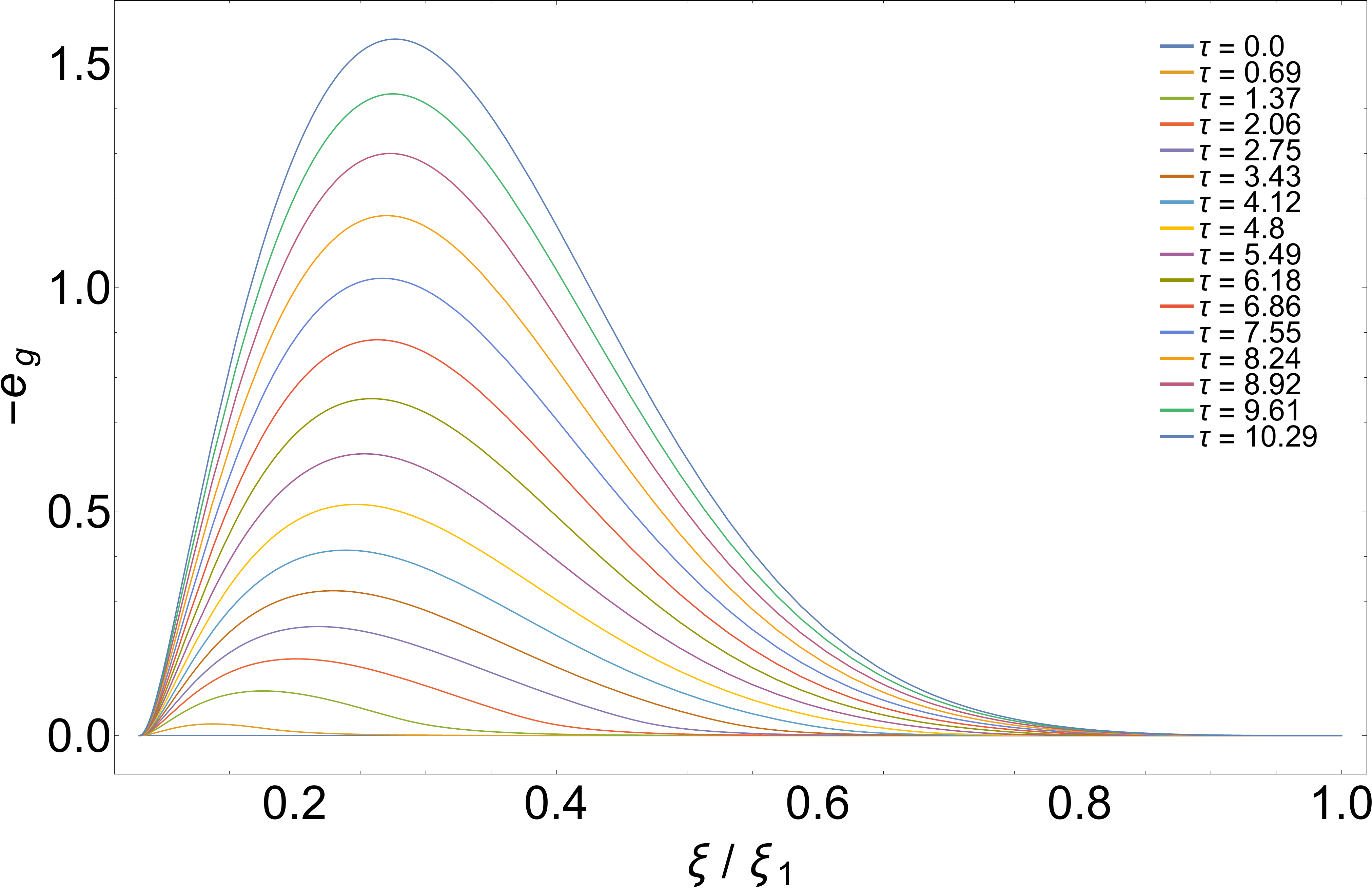}
\caption{The dimensionless kinetic energy (left panel), internal energy (middle panel), and gravitational energy (right panel) at the times shown in the legend for a $\gamma = 1.5$ polytrope, $\omega_* = 10$, and $\xi_c = 0.5$. }
\label{fig:energies}
\end{figure*}

In our polytropic model, we are effectively modeling the region exterior to where the enclosed mass is sufficient to form the neutron star and, subsequently, black hole. The free-fall time at that location is thus of order the timescale to radiate the neutrino energy (a few seconds).   We will therefore investigate the consequences of letting $\omega_*$ vary around values of order unity or larger.  Likewise, we will let the inner radius be between $\xi_c = r_c/\alpha = 0.1$ and 1 and investigate the consequences of letting it vary.  

\subsection{Velocity, mass density, and energy density}
The left-hand panel of Figure \ref{fig:velocity_radp5_omega10} shows the velocity -- normalized by the central sound speed and the total mass lost to neutrinos -- as a function of $\xi / \xi_1$ (or, equivalently, radius normalized by the radius of the progenitor), obtained from Equation \eqref{fluxex} with 600 Eigenmodes, in the envelope of a $\gamma = 1.5$ polytrope at the dimensionless times $\tau = t\sqrt{4\pi G\rho_c}$ shown in the legend. For this plot we set $\omega_* = 10$ and $\xi_c = 0.5$. The dashed, blue curve shows Equation \eqref{mach} multiplied by the dimensionless sound speed, and the black curve shows the dimensionless density profile of the polytrope scaled by a factor of 5 for clarity. The middle panel illustrates the velocity profile at a time of $\tau = 6.88$ for $\omega_* = 10$ (and is identical to the red curve in the left-hand panel; solid curve), $\omega_* = \infty$ (dot-dashed curve), and $\omega_* = 1$ (dotted curve). The right-hand panel shows the effect of changing the value of the inner radius, with the dark-red curve being the velocity profile at $\tau = 6.86$ and $\xi_c = 0.1$, while the light-red curve corresponds to $\xi_c = 1$. 

The left-hand panel of Figure \ref{fig:velocity_radp5_omega10} agrees with the general intuition established in Section \ref{sec:scalings}: at $\tau = 0$ a sound pulse is launched from the inner radius, traveling into the envelope and informing the star of the infalling core. Deep in the interior where the sound speed and density are roughly constant, the amplitude of the velocity behind the pulse does not grow substantially owing to the geometrical dilution of the energy. At later times, when the sound pulse reaches the outer extremities of the envelope where the density and sound speed go to zero, the velocity behind the sound pulse increases substantially and in approximate agreement with Equation \eqref{mach}. The middle panel of Figure \ref{fig:velocity_radp5_omega10} shows that decreasing $\omega_*$ to $\omega_* = 0.1$ (i.e, increasing the timescale over which the mass decreases) significantly decreases the amplitude of the outgoing sound pulse, because much of the star responds nearly-hydrostatically -- rather than impulsively -- to the change in mass loss.  By contrast, making the mass loss instantaneous only marginally changes the amplitude of the outgoing sound pulse. The right-hand panel illustrates that, as expected, decreasing the inner radius results in a larger velocity amplitude throughout the envelope.

The left-hand panel of Figure \ref{fig:velocity_analytic} quantifies the analytic predictions of Section \ref{sec:scalings}, and shows the early evolution of the velocity profile in the envelope of a $\gamma = 1.5$ polytrope (solid lines) when $\omega_* = \infty$ and $\xi_c = 0.5$. The dashed lines represent the dynamic scaling $v = \tau/\xi^2$ (eq. \ref{vinit}), which, at early times, we predict holds for radii outside of the radius that the sound pulse has reached in time $\tau$ -- the latter radius shown by the points in the left panel of Figure \ref{fig:velocity_analytic}  (the height of the point is the dynamic value of the velocity at that time and radius). 

Figure \ref{fig:velocity_analytic}  demonstrates that, for $\tau \lesssim 1$, the velocity profile at large radii is almost exactly equal to the simple analytic expectation of equation \ref{vinit}. The fact that the velocity changes abruptly at the sound-crossing radius also substantiates the interpretation that the boundary conditions at small radii are communicated to the fluid at the local sound speed.  In our calculation, this boundary condition is that the interior fluid elements return towards hydrostatic equilibrium while in the proto-neutron star application   the interior fluid elements would begin to collapse inwards. At later times, the analytic prediction of equation \ref{vinit} somewhat underestimates the magnitude of the velocity (though the radius at which the velocity profile peaks is still at the sound-crossing radius). This is because the differential acceleration implied by  $v \propto t/r^2$ leads to a compression of the gas, and this compression generates a pressure gradient that develops over a dynamical time and further accelerates the fluid.

The right-hand panel of Figure \ref{fig:velocity_analytic} shows the perturbation of the density induced by the mass loss for $\omega_*=10$ and $\xi_c = 0.5$ at the times in the legend (which are the same as those in Figure \ref{fig:velocity_radp5_omega10}). This figure illustrates that, while the density in the immediate vicinity of the sound pulse is slightly increased relative to the ambient value, the inner regions of the star -- and particularly those that are close to the core -- have a significantly reduced density. Thus, the sound pulse that propagates out from the core of the star is actually a \emph{rarefaction} wave in the linear theory.  This is because the reduction in the gravitational field due to the decreased central mass causes the star to expand outwards. 

Using our general expression for the flux (Equation \ref{fluxex}), the energy density \eqref{energydensity} can be written

\begin{equation}
\mathscr{E} = \frac{1}{2}\delta m^2\rho_cc_c^2\left\{e_k(\xi)+e_i(\xi)+e_g(\xi)\right\} \equiv \mathscr{E}_ce_{tot}(\xi), \label{Epoly}
\end{equation}
where $c_c^2 = \gamma p_c/\rho_c$ is the square of the sound speed at the center of the polytrope, and

\begin{equation}
e_k(\xi) = \frac{1}{\rho\xi^4}\left(\sum_{\beta}\frac{\omega_*}{\omega_*^2+\beta^2}A_\beta(\xi)R_\beta\right)^2, \label{ek}
\end{equation}
\begin{equation}
e_i(\xi) = \frac{p}{\rho^2\xi^4}\left(\sum_{\beta}\frac{\omega_*}{\omega_*^2+\beta^2}A_{\beta}'(\xi)\int_0^{\tau}R_{\beta}(\tilde{\tau})d\tilde{\tau}\right)^2, \label{ei}
\end{equation}
and
\begin{equation}
e_g(\xi) = -\frac{1}{\xi^4}\left(\sum_{\beta}\frac{\omega_*}{\omega_*^2+\beta^2}A_{\beta}(\xi)\int_0^{\tau}R_{\beta}(\tilde{\tau})d\tilde{\tau}\right)^2 \label{eg}
\end{equation}
are the dimensionless kinetic energy, internal energy, and gravitational energy, respectively. Figure \ref{fig:energies} shows the kinetic energy (left panel), internal energy (middle panel), and the absolute value of the gravitational energy (right panel) for a $\gamma = 1.5$ polytrope; we set $\omega_* = 10$ and $\xi_c = 0.5$ for these figures, and the different curves correspond to the times shown in the legend of each panel. 

From Figure \ref{fig:energies} we see that the internal and gravitational terms dominate the energetics of the inner regions of the star; this result is reasonable, seeing as the stellar envelope attempts to obtain a new, hydrostatic equilibrium in the reduced gravitational field. Immediately behind the sound pulse there is a large spike in the internal and kinetic energy, which illustrates that there is an outgoing flux of energy associated with the traveling wave. This figure also shows that, while the energy in the pulse is predominantly kinetic while it is still deep in the interior, the internal energy starts to dominate as the wave nears the stellar surface. 

\subsection{Mach number, Lagrangian positions, and shocks}
\begin{figure*}
\includegraphics[width=0.495\textwidth]{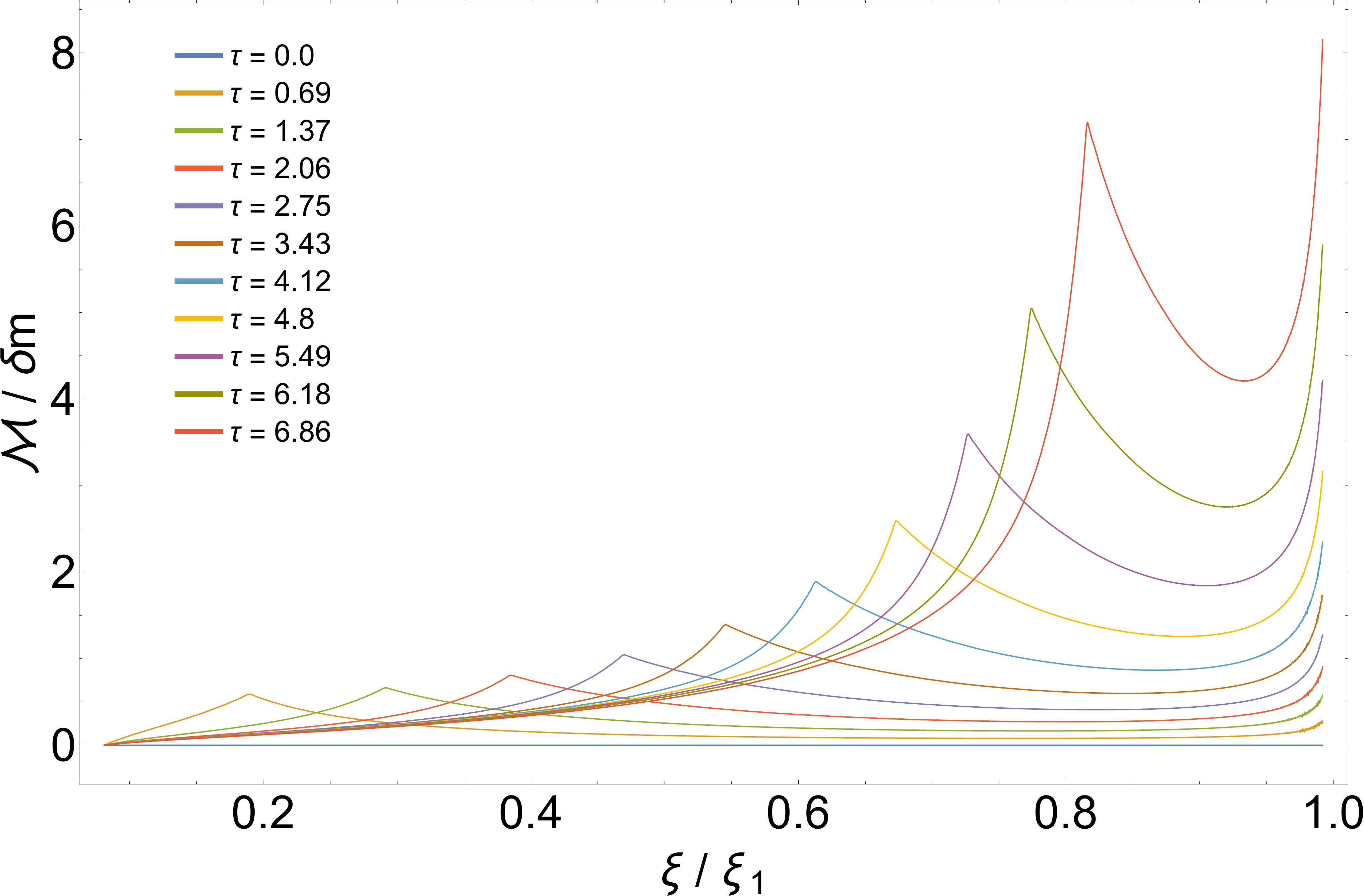}
\includegraphics[width=0.495\textwidth]{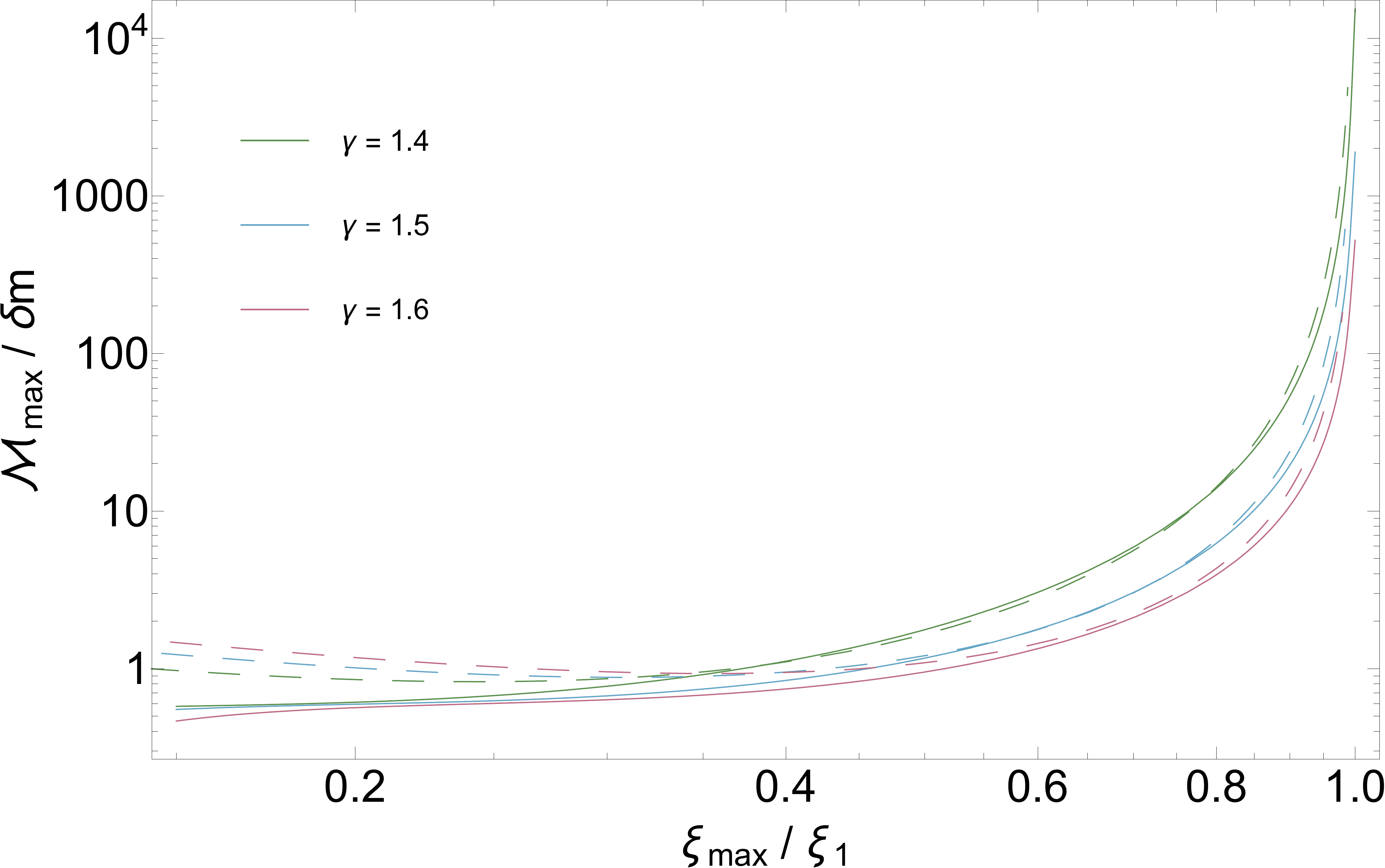}
\caption{Left: The Mach number, normalized by $\delta m$, for a $\gamma = 1.5$ polytrope, $\omega_* = \infty$, and $\xi_c = 0.5$ at the times shown in the legend. Right: The value of the Mach number immediately behind the sound pulse as a function of $\xi_{max}$ -- the location at which the maximum Mach number is reached.}
\label{fig:mach}
\end{figure*}

The left-hand panel of Figure \ref{fig:mach} shows the Mach number, normalized by the dimensionless mass loss, in the envelope of a $\gamma = 1.5$ polytrope; here we set $\omega_* = \infty$ and $\xi_c = 0.5$, and the different curves correspond to the times shown in the legend. This figure demonstrates that there are two locations -- one near the sound pulse and another near the surface -- where the Mach number becomes large. This finding is consistent with the heuristic arguments of Section \ref{sec:scalings}, where we posited that one shock should occur immediately behind the sound pulse owing to the steepening of the wave, but another could occur near the surface where the dynamic acceleration causes early (relative to the sound crossing time through the stellar envelope) shell crossing.

The location at which a shock forms in the envelope due to the steepening of the sound wave can be approximated from Equation \eqref{mach} and setting the Mach number to one. The exact value of the maximum, normalized Mach number as a function of where the maximum occurs -- both of which are functions of time as the pulse propagates into the stellar envelope -- is shown in the right-hand panel of Figure \ref{fig:mach}, where the solid curves correspond to polytropes with the polytropic index shown in the legend and we set $\omega_* = \infty$ and $\xi_c =0.5$. The dashed curves give the prediction that follows from conservation of wave power (equation \eqref{vapp}), which agrees well with the full calculation. This figure also shows that, because smaller-$\gamma$ polytropes have more extended, low-density envelopes, the shock forms sooner and at smaller radii (relative to the surface) within the star for smaller $\gamma$. 

The right-hand panel of Figure \ref{fig:mach} also gives a by-eye estimate of where the shock forms in the envelope: for physical values of $\delta m$ that are much less than one, the Mach number equals unity at a radius very near the stellar envelope. Specifically, we find for $\delta m = 0.01, 0.05, $ and 0.1 that the flow becomes supersonic at radii of $\xi = 0.97$, 0.90, and 0.85, respectively for a $\gamma = 1.5$ polytrope. In these cases, therefore, only a small amount of mass is shocked, being $\delta m_{shock} \simeq 0.1\%$, 2\%, and 8\% for $\delta m =0.01$, 0.05, and 0.1. 

In addition to the Eulerian profiles of the fluid quantities, we can investigate the Lagrangian positions of fluid shells within the stellar envelope, which are governed by the differential equation

\begin{equation}
\frac{dr_i}{dt} = v(r_{0,i},t),
\end{equation}
where $r_{0,i}$ is the initial position of fluid shell $i$. Note that, because it is already a first-order quantity, the velocity only depends on the initial position of the fluid element in this equation -- letting $v = v(r_i(t),t)$ would include higher-order terms that are not self-consistently taken into account by our perturbation approach. 

\begin{figure*}
\includegraphics[width=0.323\textwidth]{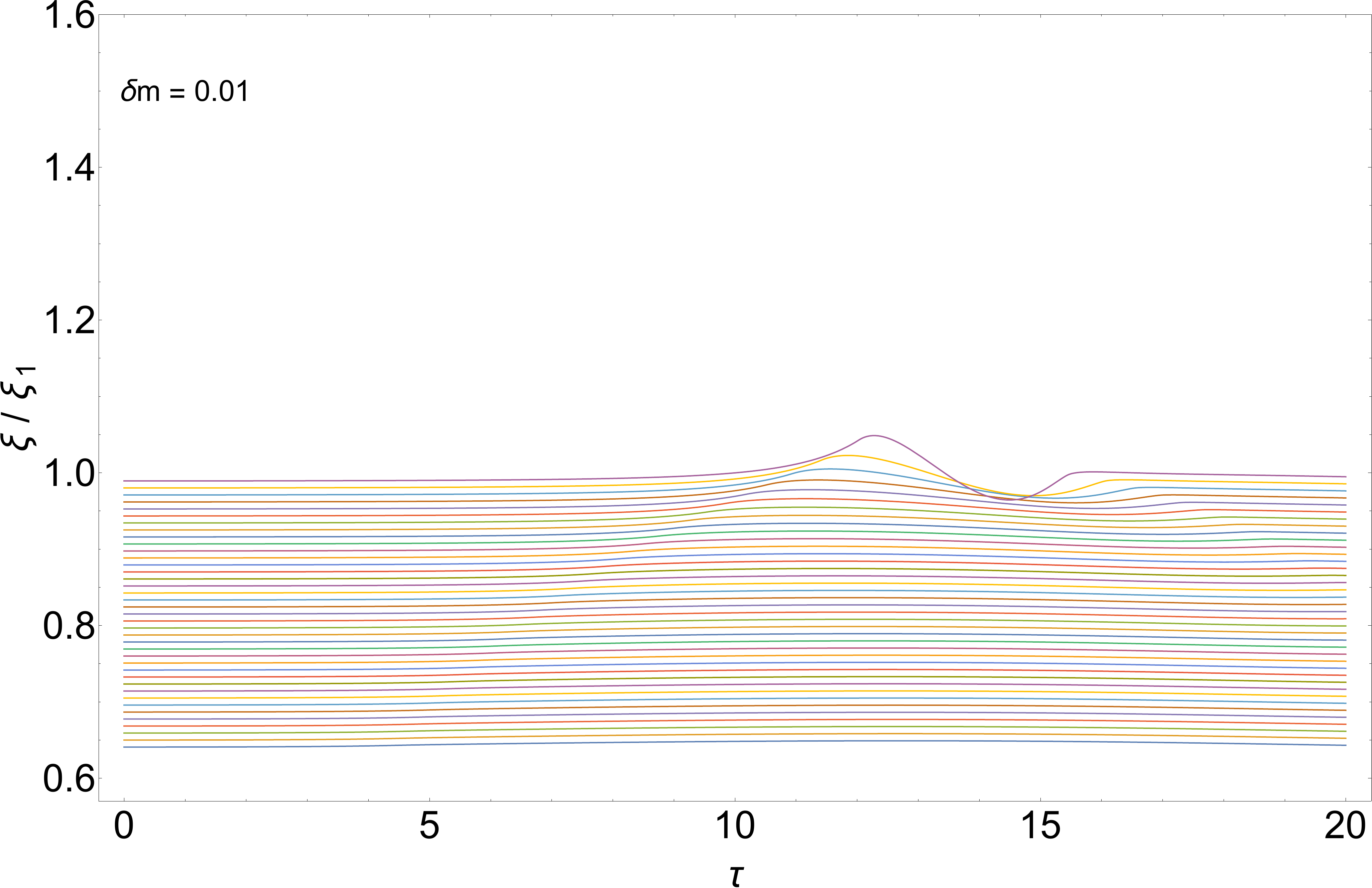}
\includegraphics[width=0.33\textwidth]{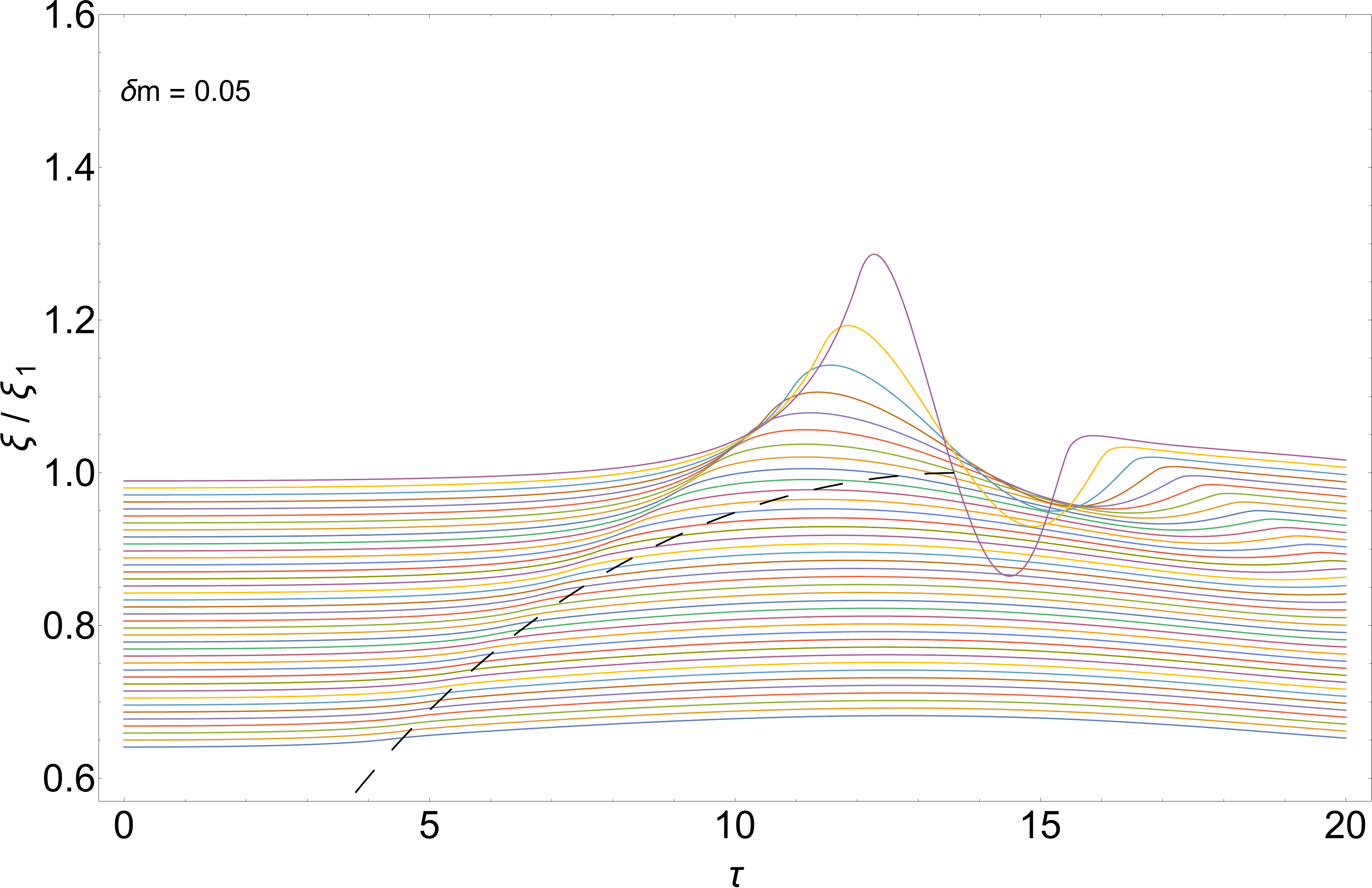}
\includegraphics[width=0.33\textwidth]{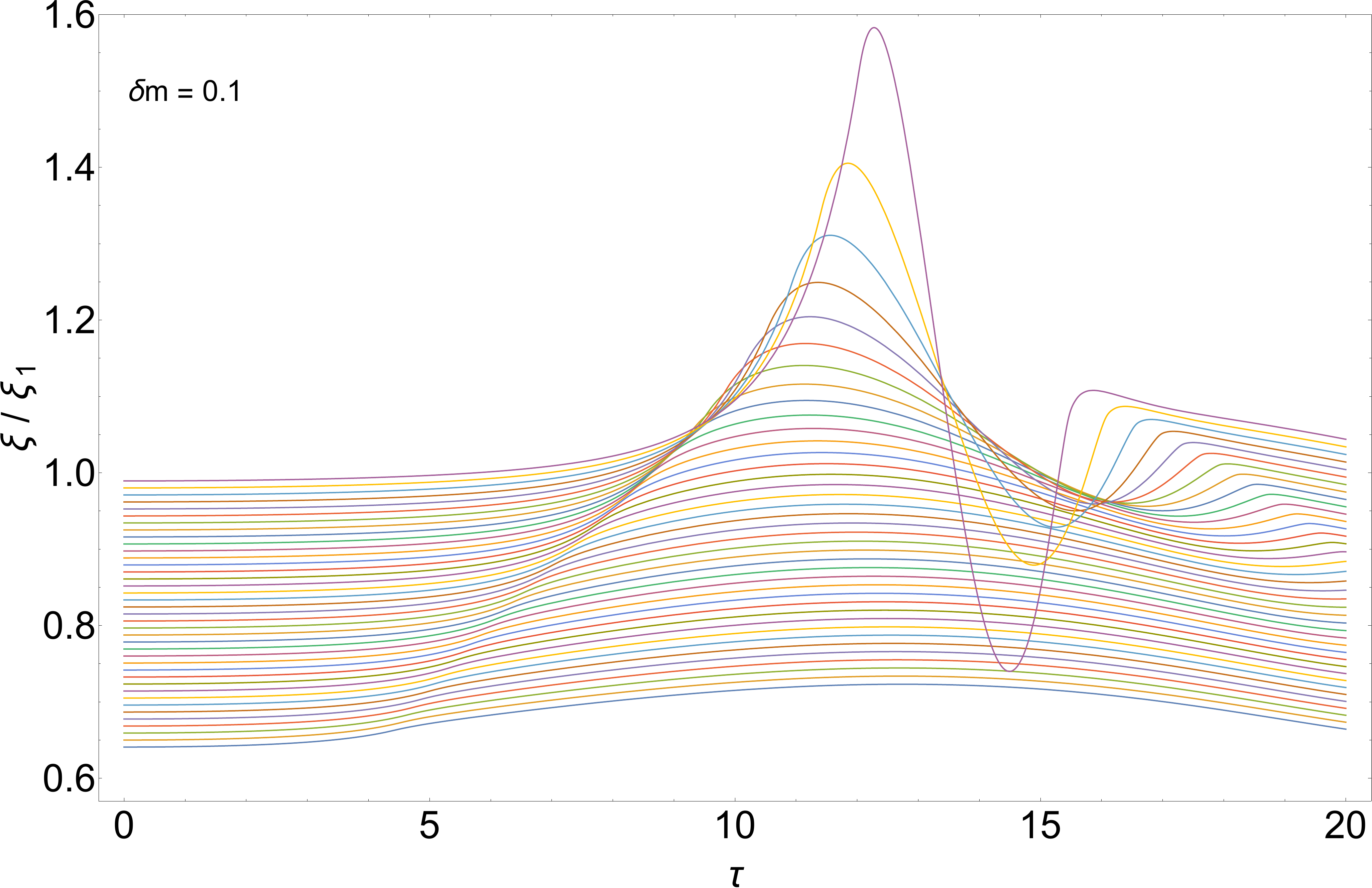}
\caption{The dimensionless, Lagrangian positions of fluid elements as a function of time within the stellar envelope of a $\gamma = 1.5$ polytrope, where different lines correspond to fluid shells with different initial positions. The three panels correspond to $\delta{m} = 0.01$ (left), $\delta{m} =0.05$ (middle), and $\delta{m} = 0.1$ (right), and we set $\omega_* = \infty$ and $\xi_c = 0.5$ for these panels. The black, dashed line in the middle panel shows the position of the sound pulse as a function of time.  Shell crossings indicate where shocks would form in a nonlinear treatment.  Note that these occur in two places:  at the surface prior to the sound pulses arrival and as the sound wave reaches large amplitudes in the low density surface layers of the star.}
\label{fig:lags}
\end{figure*}

Figure \ref{fig:lags} illustrates the Lagrangian positions of the fluid shells within the stellar envelope of a $\gamma = 1.5$ polytrope, where we set $\omega_* = \infty$ and $\xi_c = 0.5$. The different lines correspond to different initial positions in the progenitor, and each panel has a total mass lost indicated in the upper-left corner of the plot. The dashed, black line in the middle panel shows the position of the traveling sound wave as a function of time. We see that, for very small values of the mass loss, the positions of the fluid elements are only slightly perturbed from their initial positions, and small-amplitude oscillations are excited in the outer part of the star. However, as $\delta m$ increases, fluid elements are increasingly displaced, and this is especially true for mass shells near the surface of the star where the oscillations become particularly intense. 

We also see from this figure that fluid elements cross after being ``hit'' by the sound pulse, as these shells at large scale heights in the atmosphere attempt to return to a new equilibrium and cross in the process. This shock is thus caused by the steepening of the sound pulse as it propagates into the more rarefied atmosphere of the star. There is also a second shock that forms as shells very near the surface cross prior to being reached by the sound pulse. This shock is independent of the steepening of the pressure wave, and is related to the differential acceleration that results from the initial, dynamic response of the envelope to the gravitational field. Consistent with Figure \ref{fig:mach} and the discussion in Section \ref{sec:scalings}, we therefore see that there are two locations at which shocks form in the polytrope -- one due to the steepening of the sound wave, and another from the differential acceleration caused by the dynamic expansion.

\subsection{Total energy}
From Figure \ref{fig:energies} we see that the internal energy of the inner regions of the envelope -- through which the sound pulse has already passed -- has been augmented significantly. This additional energy results from the fact that, as the sound wave passes through the envelope, the star attempts to reconfigure itself into a new, hydrostatic equilibrium in the reduced gravitational field. Thus, the kinetic energy originally contained in the gas is transferred into internal energy.

However, it is also apparent from Figure \ref{fig:energies} that the gas in the immediate vicinity of the sound pulse has a very localized increase in energy. The precise definition of this ``sound pulse'' and its associated energy is somewhat arbitrary, but we define the energy in the sound wave as the total, integrated energy contained within the full-width at half-maximum around the peak in the velocity profile. The precise limits on velocity used to define the energy  do not significantly affect our results,  especially  once the sound pulse nears the surface of the star and the velocity profile becomes increasingly peaked. This definition also encapsulates the physical idea that it is only  fluid elements  moving along with the sound pulse that would ultimately pass through the shock formed when the sound wave steepens.

\begin{figure}
\includegraphics[width=0.47\textwidth]{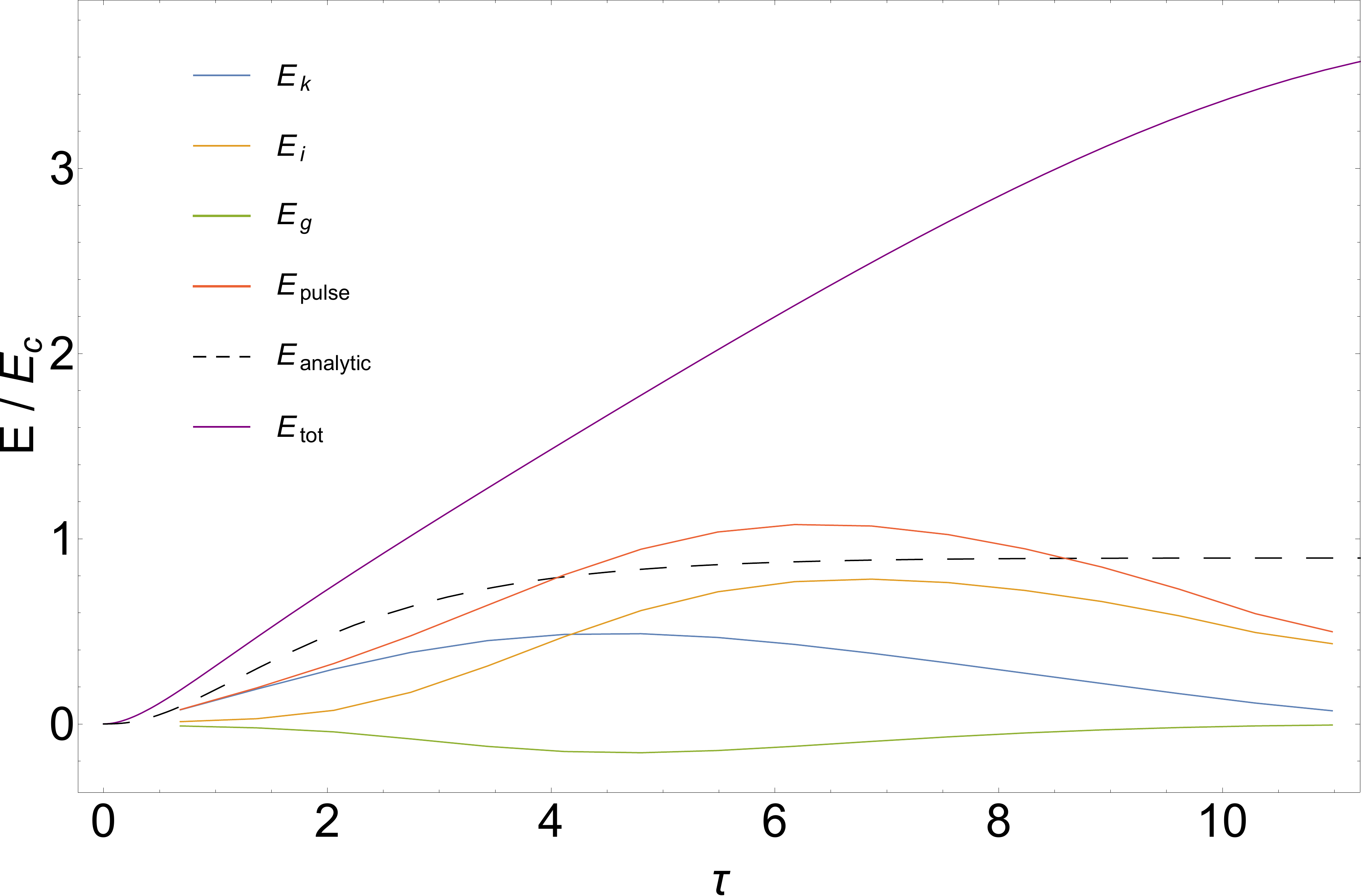}
\caption{The kinetic energy (blue curve), internal energy (orange curve), gravitational energy (green curve), and the sum of these energies (red curve) contained in the outgoing sound pulse as a function of time in a $\gamma = 1.5$ polytrope with $\omega_* = \infty$ and $\xi_c = 0.5$; the individual energies follow from the integrals of Equations \eqref{ek}, \eqref{ei}, and \eqref{eg} over the full-width at half max of the velocity around the peak in the pulse. The black, dashed curve shows the analytic approximation that follows from Equation \eqref{Eapp}, and the purple curve gives the total, integrated energy in the star (from Equation \ref{Etotal}).} 
\label{fig:energies_gamma1p5}
\end{figure}

\begin{figure*}
\includegraphics[width=0.495\textwidth]{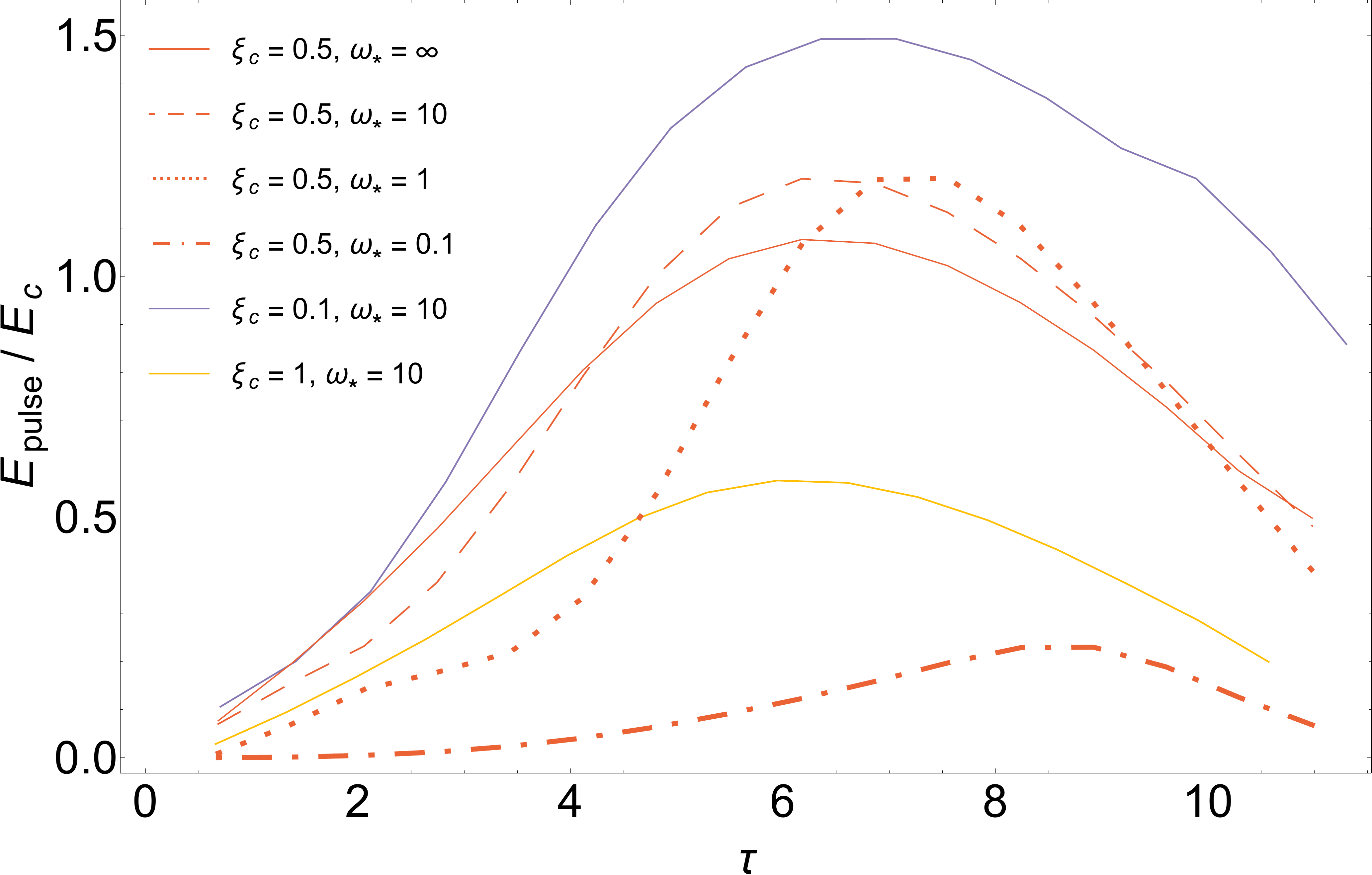}
\includegraphics[width=0.495\textwidth]{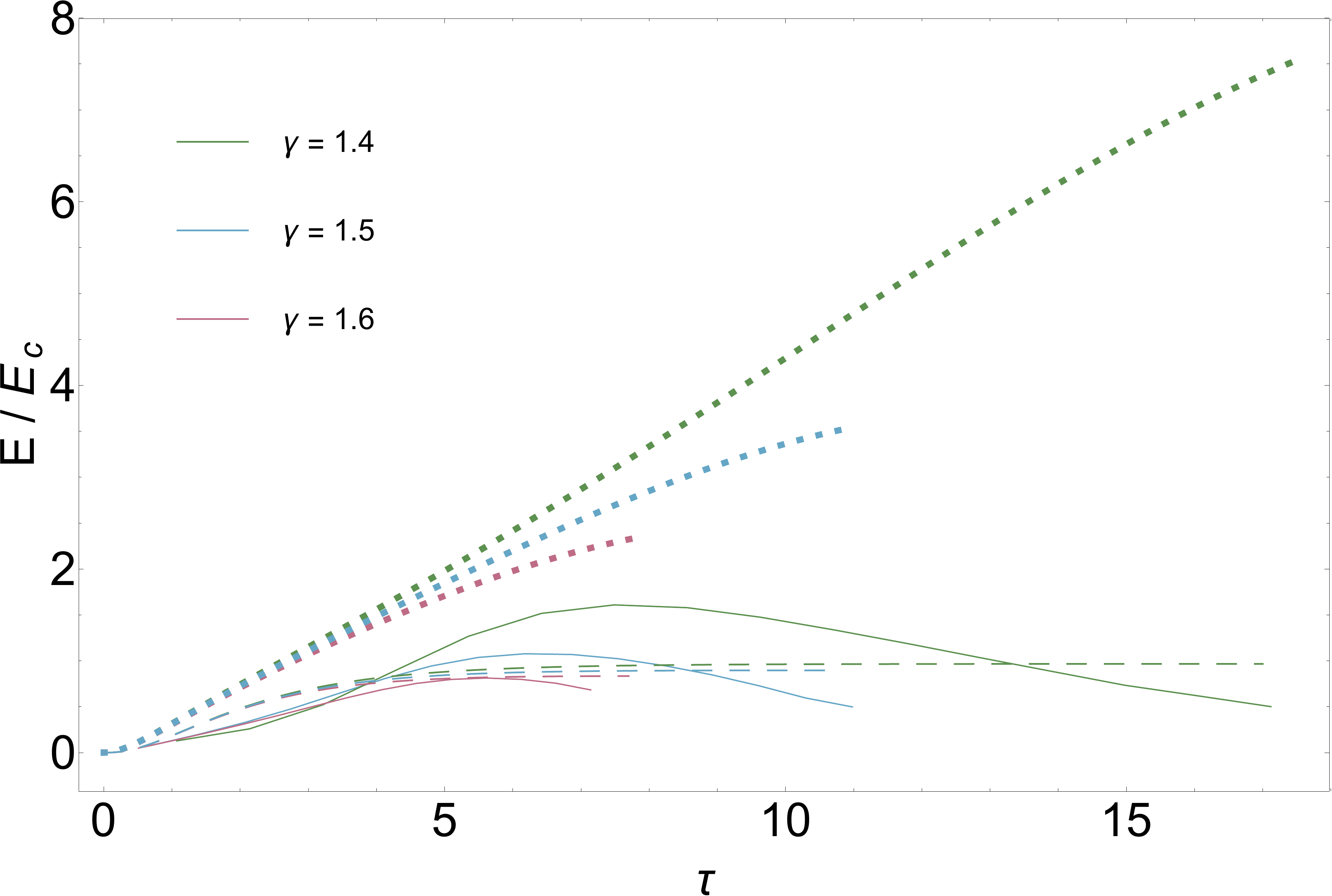}
\caption{Left: The total, integrated energy over the sound pulse for a $\gamma = 1.5$ polytrope, where the different curves correspond to the different parameters (i.e., $\omega_*$ and $\xi_c$) shown in the legend. Right: The integrated energy in the sound pulse (solid curves), the analytic approximation of the energy in the sound pulse (dashed curves), and the total energy contained in the star (dotted curves) as functions of time for $\xi_c = 0.5$ and $\omega_* = \infty$ and the polytropes shown in the legend.}
\label{fig:E_pulse_comps}
\end{figure*}
Figure \ref{fig:energies_gamma1p5} shows the kinetic ($E_k$), internal ($E_i$), and gravitational ($E_g$) energies contained in the sound pulse, which are given by the integrals of Equations \eqref{ek}, \eqref{ei}, and \eqref{eg} over the width of the sound pulse, and the sum of these energies ($E_{pulse}$) as functions of time. These curves are for a $\gamma = 1.5$ polytrope with $\omega_* = \infty$ and $\xi_c = 0.5$. The analytic approximation to the energy introduced in Section \ref{sec:scalings} -- specifically the first equality in Equation \eqref{Eapp} -- is shown by the black, dashed curve, and the total energy contained in the star (Equation \ref{Etotal}) is given by the purple curve. 

Figure \ref{fig:energies_gamma1p5} demonstrates that, while the energy contained in the pulse is not exactly constant in time (indeed, it must be zero at $\tau = 0$ when the perturbations are everywhere zero), the late-time behavior -- once the pulse has a well-defined peak in velocity in the outer extremities of the envelope -- of the energy is approximately constant around $E \simeq E_c$. We also see that the analytic expression of Equation \eqref{Eapp} does a very good job of approximating both the increase in the energy of the pulse and the average value at later times. 

The total energy change in the star induced by the mass loss overestimates the energy contained in the outgoing pulse by a factor of roughly three. This overestimate is due to the fact that the velocity increases non-locally within the star from the initial, dynamic response, and this additional kinetic energy is not captured in the integral over the FWHM of the pulse. Furthermore, the inner regions of the star are significantly overpressured, and this increase in the internal energy (which is a relic of the passage of the sound wave) is not contained in the outgoing sound wave. 

The left-hand panel of Figure \ref{fig:E_pulse_comps} shows how the integrated energy contained in the sound pulse generated in the envelope of a $\gamma = 1.5$ polytrope varies as a function of $\xi_c$ and $\omega_*$, the various curves corresponding to the values of these parameters shown in the legend. We see that as long as $\omega_* \gtrsim 1$  -- which we expect based on physical grounds -- changing the timescale over which the mass is radiated does not significantly affect the total energy contained in the outgoing wave (though the growth of the energy is slightly impeded for smaller $\omega_*$). Similarly, decreasing (increasing) the value of the inner radius only marginally increases (decreases) the energy contained in the outgoing pulse. 

The right-hand panel of Figure \ref{fig:E_pulse_comps} shows the variation in the mass-loss-induced energy for different polytropes, given by the polytropic indices in the legend, when $\xi_c = 0.5$ and $\omega_* = \infty$. In this figure, the solid lines represent in the energy integrated over the width of the sound pulse, the dashed lines are the analytic approximation from Section \ref{sec:scalings}, and the dotted curves are the total energy contained in the star (from Equation \ref{Etotal}). Interestingly, while the analytic estimates and the energy in the sound pulse are all fairly similar, the total energy contained in the star increases substantially for stars with smaller $\gamma$ (and more extended envelopes), and the change in the energy has a more dramatic effect on the restructuring of the stellar interior in the reduced gravitational field. Given that $\gamma=4/3$ polytropes have zero total energy \citep{hansen04}, this result is reasonable: as $\gamma \rightarrow 4/3$, the relative change in the energy due to the mass loss becomes more severe, and the final state of the star is more significantly perturbed from its initial, hydrostatic one.

\section{Application to Real Stellar Progenitors}
\label{sec:stars}
We saw in the previous section that polytropes develop many of the features exhibited by the simulations of \citet{nadezhin80,lovegrove13,lovegrove17,fernandez17}: the mass loss in the core generates an outward-propagating pressure wave that traverses the star. As it propagates down the density gradient of the stellar envelope, the sound wave grows in amplitude, and in a fully nonlinear treatment would form a shock. This shock will subsequently erupt from the surface in a relatively mild explosion, taking with it some portion of the stellar envelope.

While this qualitative agreement is encouraging, some questions remain unanswered. For example, the estimate of the energy given by Equation \eqref{Eapp} -- which was substantiated by the polytrope models -- relies on specifying the pressure scale height $\alpha$. For a polytrope, this is uniquely determined by the central pressure and temperature; however, the iron core (the true, geometric center) of a real star is orders of magnitude denser than the helium envelope, and using the central scale height would give a much larger estimate of the energy than would be obtained from, for example, the average scale height throughout the envelope. Furthermore, real stars are not necessarily well-represented by polytropes, and estimates of where the shock forms will differ from those found in Section \ref{sec:polytrope}. 

In this section we will use our previous results to investigate the properties of shocks formed by neutrino energy radiation in real stars,  making specific comparisons to the models analyzed in F17.  For these purposes we use some of the properties of the pre-collapse stellar progenitors from F17 calculated using MESA \citep{paxton11,paxton13,paxton15}.  Table \ref{tab:mesa} summarizes some of the key properties of these models, along with of the predictions of our analytic results in this paper.

\begin{table*}
\begin{center}
{\bfseries Properties of MESA Pre-Collapse Models}
\end{center}
\begin{center}
\begin{tabular}{lcccccccc}
\hline
 $M_{ZAMS}$  & $M_{cc}$ & $R_{cc}$ & Type & $\xi_{2.5}$ & $r_c$ & $\mathcal{M}_{surf}$ & $r_{sh}$ & $E_{pulse}$ \\
  ($M_\odot$) & ($M_\odot$) & ($R_\odot$) & & & ($R_\odot$) & & $R_{\odot}$ & $10^{48}$ erg \\
\hline
15 &  10.8 & 1060 & RSG & 0.24 & 0.02 & 0.3 & 2.1 & 1.5\\
25 & 11.7 & 96 & BSG & 0.33 & 0.024 & 0.3 & 0.37 & 1.2\\
40 & 10.3 & 0.38 & WR & 0.37 & 0.02 & 1.7 & 0.22 & 1.7\\
\hline
\end{tabular}
\end{center}
\caption{The properties of the three, fiducial solar metallicity {\sc mesa} models analyzed in F17:  the ZAMS mass, the mass of the star at the onset of core collapse, the radius of the star at the onset of core collapse, the type of star, the compactness, and the inner radius where we expect the energy to be generated in the outgoing sound pulse. The last three columns give the analytically-predicted upper limit to the surface Mach number at the time the sound pulse reaches the stellar photosphere, the radius at which the sound pulse steepens into a shock, and the energy contained in the sound pulse.} \label{tab:mesa}
\end{table*}

\subsection{Sound Pulse Energetics}
In a real star, collapse is initiated by the cessation of nuclear burning in the iron core, and the subsequent de-leptonization and neutrino radiation. The resulting loss of pressure support  in the interior causes successive shells of the star to infall onto the newly forming proto neutron star. The effective inner radius $r_c$  in equation \eqref{Eapp} for a real stellar progenitor depends on the structure of the progenitor, in particular via the compactness parameter\footnote{We utilize the standard notation for compactness but note the possible confusion with the dimensionless polytropic radius from the previous section.  Context should make it clear which is which.}
\begin{equation}
\xi_{2.5} = \frac{2.5}{r(M_r=2.5 M_\odot)/10^8 \, {\rm cm}}.
\label{xi}
\end{equation}
Note that the free-fall time at the 2.5 $M_\odot$ mass coordinate can be written as $t_{ff}(M_r = 2.5 M_\odot) \simeq 0.2 \xi_{2.5}^{-3/2}$ sec.   For high compactness $\xi_{2.5} \gtrsim 0.2$, the free-fall time at the mass coordinate of the maximum mass of a neutron star is less than the few seconds characterizing neutrino diffusion out of the proto-neutron star \citep{burrows88}. In this case the neutron star is formed, radiates only a fraction of its binding energy in neutrinos, and collapses to a black hole on the infall timescale at $r_c \simeq 2.5 \times 10^8 \, \xi_{2.5}^{-1}$ cm, and hence $\omega_* = \omega \times t_{ff} \simeq 1$ in the notation of the previous sections, where $t_{ff}$ is the free-fall time from $r_c$. On the other hand, if the progenitor is not very compact, with $\xi_{2.5} \lesssim 0.2$ (and yet for some reason still does not successfully explode), the radius enclosing $\sim 2.5 M_{\odot}$ has a long free-fall time relative to the neutrino diffusion time.   In this case most of the neutrino radiation occurs prior to the collapse to a black hole and $r_c$ is  set by where the free-fall time is comparable to the neutrino diffusion time of a few seconds, and $\omega_*$ is again effectively $\sim 1$ given this value of $r_c$.   To bracket both of these regimes, we can define $t_{ff}(r_c) \simeq {\rm min}(\tau_c,\tau_{tov})$, where $\tau_c \sim$ few sec is the neutrino cooling time of a proto-neutron star and $\tau_{tov}$ is the time for the proto-neutron star to collapse to a BH.   The corresponding values of $r_c$ are given in Table \ref{tab:mesa}.  Note that $r_c \sim 0.02 R_\odot \sim 1.5 \times 10^9$ cm for most progenitors.  This is true except for the most massive ones with high compactness (not shown in Table \ref{tab:mesa}), for which the time to form a BH is short, suppressing neutrino radiation.   In these models the mass radiated in neutrinos $\delta M$ will also be correspondingly smaller.   

\begin{figure}
\includegraphics[width=0.5\textwidth]{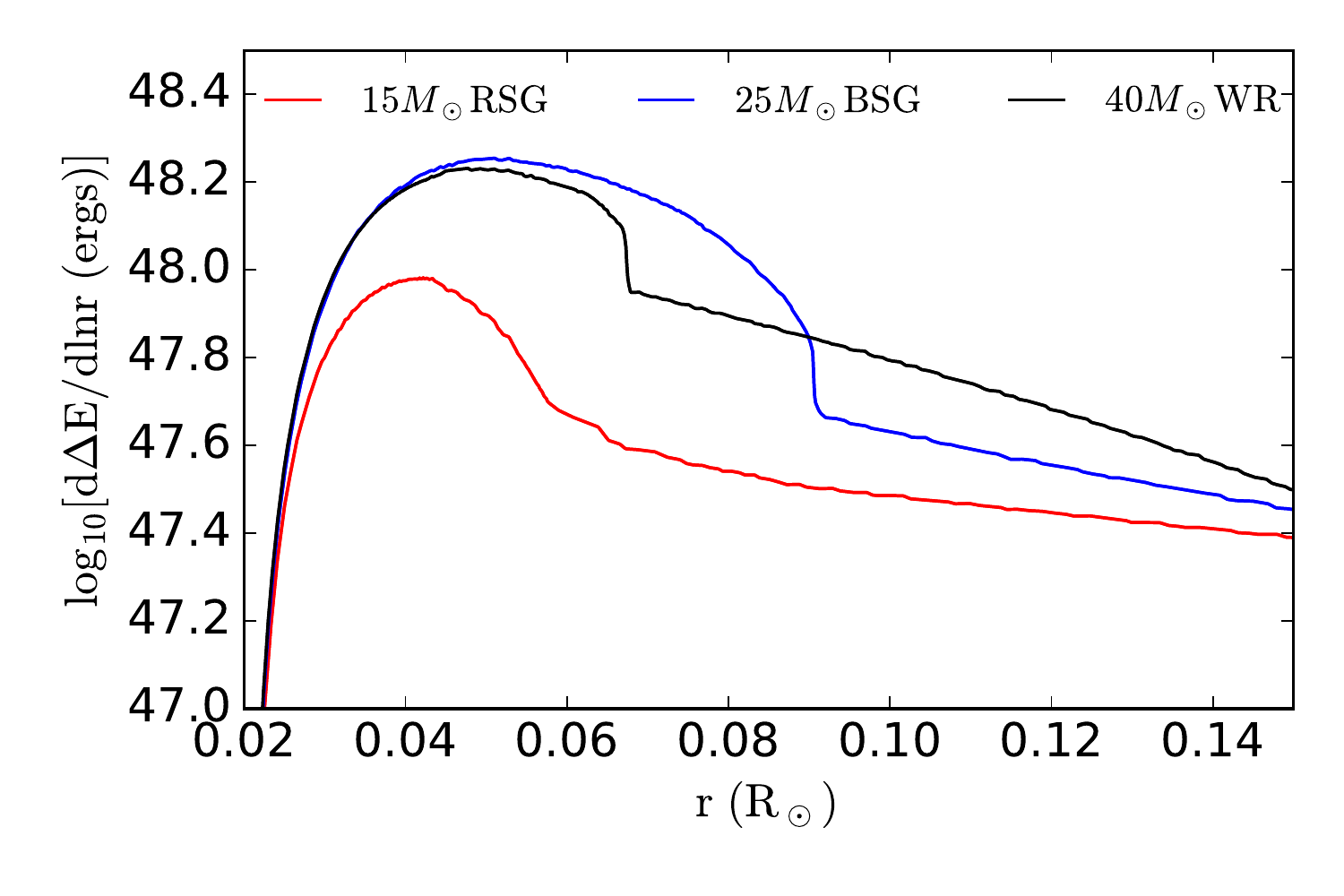}
\caption{Analytic estimate (eq. \ref{eq:dEloc}) of energy input to the sound pulse as a function of radius, produced by neutrino radiation of $\delta M = 0.3 M_\odot$. The progenitors are solar metallicity MESA models (Table \ref{tab:mesa}).   The energy input is suppressed interior to $\sim 0.02 R_\odot$ because the mass collapses to form the proto-neutron star.}
\label{fig:E_pulse_progenitors}
\end{figure}

When $\omega_* \gtrsim 1$, the total energy contained in the outgoing sound pulse (that eventually steepens into the shock) is not strongly affected by $\omega_*$, as shown by the left-hand panel of Figure \ref{fig:E_pulse_comps}. The analytic prediction of Section \ref{sec:scalings}, which agrees well with the more exact, perturbation analysis (the right-hand panel of Figure \ref{fig:E_pulse_comps}), then gives an energy injection at a given radius of
\begin{equation}
\frac{d \Delta E}{d \ln r} \simeq \frac{2 \pi \rho G^2 \delta M^2 \tau_{sc}^2}{r}
\label{eq:dEloc}
\end{equation}
where $\tau_{sc}$ is given by equation \ref{eq:tsc}.

Figure \ref{fig:E_pulse_progenitors} shows the estimated energy input for solar metallicity MESA progenitors as a function of radius, for $r_c = 0.02 R_\odot$ and $\delta M = 0.3 M_\odot$.  Table \ref{tab:mesa} gives the total energy $E_{pulse}$ integrated over all radii.  The result is $\sim 10^{48}$ erg. These estimates are in good agreement with the sound pulse energy in the interior in the full simulations of F17 (see their Fig. 5).   We note, however, that the final shock energy erupting from the surface can be significantly lower in some cases due to energy lost as the shock propagates through the outer layers of the star.   We defer a theoretical analysis of that phase to future work.    

Finally, we note that the amount of mass that can be ejected given the energy scale shown in Figure \ref{fig:E_pulse_progenitors} varies significantly with progenitor.  For RSGs, BSGs, and WR stars it is $\sim 5$, 0.1, and $10^{-3} \, M_\odot$, respectively, due to the increasing binding energy of the envelope for more compact progenitors.

\subsection{Where Does the Sound Pulse Become a Shock?}
From the analysis of Section \ref{sec:polytrope}, the transition from the linear sound pulse into a shock occurs near the surface where the density and sound speed drop considerably. For the polytropic models, the linear theory prediction for the location where the pulse becomes supersonic was found by using the general expression for the mass flux in the star (eq. \ref{fluxex}). However, conservation of the energy flux (wave power) associated with the sound pulse also  gives approximately the same results, as shown by the right-hand panel of Figure \ref{fig:mach}. 

The stellar models presented in F17 range from Wolf-Rayets to red supergiants, and thus have a wide variety of envelope properties. Their simulations demonstrate that the strongest shocks (in terms of the maximum Mach number reached and the amount of mass that experienced the shock) are formed when $r_c$ is a small fraction of the stellar radius, which agrees qualitatively with our analytic findings: in these instances, the energy is injected at a relatively small radius and the Mach number has a long time to grow as the pulse propagates down the density gradient. On the other hand, the Wolf-Rayet progenitors do not generate shocks until very near the stellar surface, which is consistent with the fact that $R_{cc}/r_c \sim 10$ in those cases -- the sound wave only significantly steepens as the density declines drastically near the photosphere. 

\begin{figure}
\includegraphics[width=0.47\textwidth]{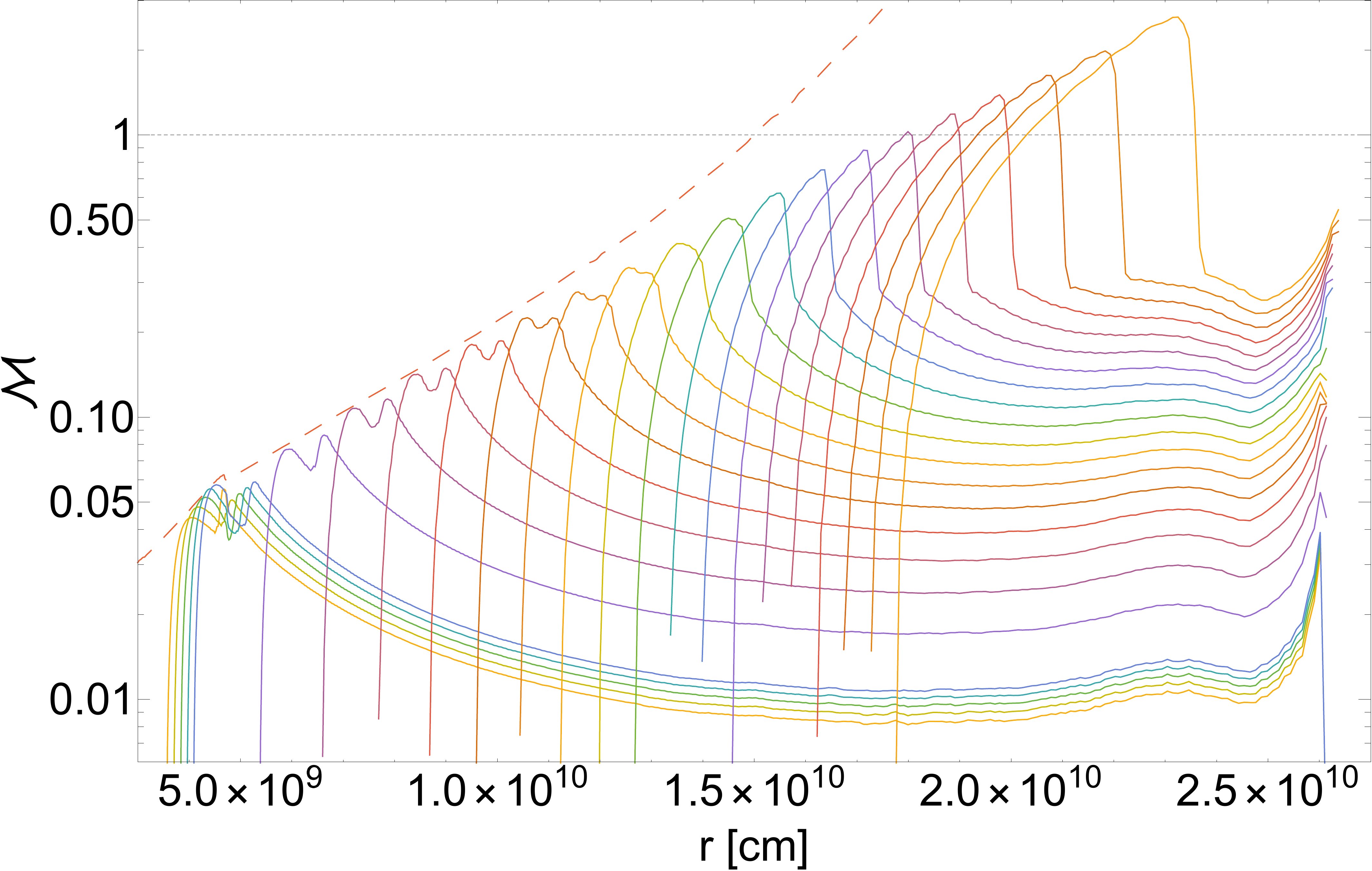}
\caption{Solid lines show the Mach number at different times from the 40 $M_{\odot}$, WR progenitor analyzed in F17. The red, dashed curve shows the analytic prediction (equation \ref{mach}) which accurately describes both the increase in amplitude of the sound pulse and the radius at which $\mathcal{M} \sim 1$, where the shock first forms.}
\label{fig:mach_m40_comp}
\end{figure}

Figure \ref{fig:mach_m40_comp} shows the Mach number as a function of radius $r$ from the simulation of a 40 $M_{\odot}$ Wolf-Rayet star analyzed in F17. The solid lines show the results of the full numerical simulations at different times (ranging from 0 to $\sim$ 100 s post-mass-loss), while the red, dashed curve gives the analytic prediction from Equation \eqref{mach} with the normalization chosen to match the numerical scaling at radii $\sim r_c$.  Alternatively, directly applying equation \eqref{mach} with $\delta M/M_{in} = 0.058$ (which corresponds to the values in the simulation of F17) yields a similar normalization if the density, radius, and sound speed in the square root in equation \eqref{mach} are normalized at radii $\sim 2-3 r_c$, which is where the energy input peaks (see Fig. \ref{fig:E_pulse_progenitors}).  This reflects the fact that linear theory accurately predicts both the energy and length scales associated with the formation of the sound pulse.   Figure \ref{fig:mach_m40_comp}  demonstrates that the increase of the Mach number throughout the star in the linear phase is well-matched by the conservation of the power of the sound pulse.   The radius at which the shock forms predicted by linear theory is accurate to about $\sim 50 \%$.

\subsection{Does the Photosphere Shock Prior to the Sound Pulses Arrival?}
As described in Section \ref{sec:scalings} and \ref{sec:polytrope}, it is in principle possible that the surface of a star feels two shocks associated with the response to the neutrino radiation, the first as the cool surface layers accelerate according to equation \ref{vinit}, and the second when the sound pulse from the interior reaches the surface. This double shock occurs for polytropes because the sound speed goes to zero at the surface; indeed, this is the defining quality of the surface of a polytrope.   To assess whether this occurs for real stellar progenitors -- where the surface coincides with the photosphere and a non-zero sound speed -- we estimate the surface Mach number using equation \ref{vinit} with $t$ set by the global sound crossing time of the star
\begin{equation}
\mathcal{M}_{surf} \lesssim \frac{G \delta M}{R^2 c_s(R)} \int_{r_c}^R \frac{dr}{c_s}.
\label{eq:Msurf}
\end{equation}
Equation \ref{eq:Msurf} is an upper limit (by a factor of few)  because the sound pulse steepens into a shock that travels supersonically through the outer part of the stellar envelope, thus reaching the surface on a time-scale somewhat shorter than $\tau_{sc}$.

Table \ref{tab:mesa} gives our estimates of $\mathcal{M}_{surf}$ for the MESA progenitors, for $\delta M = 0.3 M_\odot$ and $c_s(R)$ evaluated at the photosphere.  We see that $\mathcal{M}_{surf} \sim 0.3$ for the RSG and BSG but $\mathcal{M}_{surf} \sim 1$ for the WR model, implying that the initial dynamical acceleration of the surface is significant even prior to the sound pulse arrival.  We find similar values for other WR models and that in general the compact progenitors are the most likely to have high photospheric velocities in the dynamical acceleration phase.  The exception to this is very compact massive progenitors with large compactness, which form black holes so quickly that the mass radiated in neutrinos is significantly smaller than $\delta M \sim 0.3 M_\odot$.

Figure \ref{fig:mach_m40_comp} shows that in F17's simulations of a 40 $M_\odot$ WR progenitor, the maximum Mach number of the photosphere prior to the shock reaching it is $\sim 0.5$, reasonably consistent with our analytic conclusions here.  We note that the photosphere is not fully resolved in F17's simulation because of the small surface scale-height for WR stars so it is likely that the true surface Mach number is somewhat higher.    Assessing whether the photosphere shocks, modifying the observed emission, will require additional simulations with resolution focused on the photosphere and likely the inclusion of the WR star's wind since in WR stars the photosphere is often out in the stellar wind. For comparison, for F17's 15 $M_\odot$ RSG progenitor, for which the photosphere is better resolved, the simulation yields a surface Mach number of $\sim 0.2$ prior to the shock reaching the photosphere, very close to our analytic prediction. In future work it would be interesting to calculate how the initial dynamical acceleration of the photosphere imprints itself on the emission even prior to the shocks arrival at the surface.

\section{Summary and Conclusions}
\label{sec:conclusions}
During the formation of the protoneutron star in the core-collapse of a massive star, the emission of neutrinos results in a decrease in the mass of the core by $\sim 0.3 M_{\odot}$.  If the accretion shock onto the protoneutron star fails to revive, leading ultimately to the collapse of the star to black hole, the decrease in  gravitational acceleration caused by neutrino mass loss still produces both a bulk outward motion of the entire outer stellar envelope as well as a pressure wave that propagates through the star and steepens into a shock near the stellar surface. This effect, first analyzed by \citet{nadezhin80}, is a critical part of the disappearance of a star in an otherwise-failed supernova.

In this paper we developed a general formalism for understanding the physical origin of this sound pulse, its energetics, and its propagation through the star.    We argued (Section \ref{sec:scalings}) heuristically that the energy contained in the outgoing wave should be $E_{pulse} \simeq 10^{48}$ erg, relatively independent of progenitor. This estimate agrees well with the results of simulations (\citealt{lovegrove13,fernandez17} (F17 in this paper)). We then exploited the fact that the fractional change in mass of the star produced by neutrino mass loss is small, $\lesssim 0.1$.   This means that the initial excitation and propagation of the sound pulse can be accurately calculated using linear perturbation theory.  We do so by writing the velocity and density profiles of the evolving stellar envelope in terms of the Eigenmodes of the unperturbed star (Section \ref{sec:general}).  The linear analysis eventually breaks down as the sound pulse grows in amplitude propagating into the lower density stellar envelope.   We can, however, accurately predict where the resulting shock forms using our linear theory results.

In Section \ref{sec:polytrope}, we applied the results of our perturbation analysis to polytropic stellar models.  Many of the resulting features are in good qualitative agreement with simulations: a sound pulse is launched from the inner boundary, growing in amplitude and eventually becoming supersonic near the stellar surface. The conservation of the power of the wave $L_{wave}\simeq 4 \pi \rho r^2 v^2 c_s$ gives an excellent prediction of the growth of the Mach number throughout the envelope (right-hand panel of Figure \ref{fig:mach}).  We then used this agreement to investigate where the sound pulse will form a shock in more realistic stellar progenitors (Section \ref{sec:stars}), in particular the Wolf-Rayet, red supergiant, and blue supergiant {\sc mesa} progenitors analyzed numerically in F17. Our analytic predictions of the radius where the sound pulse is excited, the initial energy in the sound pulse, and the radius where the shock first forms agree well with the numerical simulations of F17 (see, e.g., Figs. \ref{fig:E_pulse_progenitors} \& \ref{fig:mach_m40_comp}).

Our analytic results also demonstrate that, in principle, a shock can form in the stellar photosphere {\em prior} to the sound pulses arrival.  This is a consequence of the initial, dynamic expansion of the star prior to the passage of the sound wave, and the low sound speed near the stellar photosphere.  This shock does inevitably form in polytropic models for which the sound speed vanishes at the surface (Fig. \ref{fig:mach} \& \ref{fig:lags}).  For realistic stellar progenitors we find that the photospheric Mach number is limited to $\sim 0.2$ (RSG, BSG) and $\sim 1$ (WR) (see, e.g., equation \ref{eq:Msurf} and Figure \ref{fig:mach_m40_comp}).  This expansion of the stellar photosphere is likely to slightly decrease the stellar effective temperature prior to the weak shock breakout and the star's subsequent disappearance.   In some WR stars the photosphere may actually undergo `internal shocks' via this process but more detailed calculations are required to assess this, including the fact that in WR stars the photosphere is often out in the wind (see Section \ref{sec:stars}).  

Our linear treatment is useful for predicting where in the star the shock forms and the amount of energy it contains as it does so. Therefore, one can use our results as a starting point for the further investigation of the shock propagation through the remainder of the stellar envelope. The shock energy is much less than in typical core-collapse supernovae and the Mach number is only of order unity.  As a result, existing theoretical calculations of shock propagation, which focus on strong shocks and ignore the gravitational energy (e.g., \citealt{matzner99}), may not be applicable to these very low energy shocks. Indeed, the simulations of \citet{fernandez17}, which follow the eventual emergence of the shock from the photosphere of their progenitors, find that the shock can in some cases lose a significant amount of energy before emerging from the surface. It is unclear whether this is primarily energy lost due to the entropy trail left behind by the shock or the effects of gravity on the shock propagation.   It would be valuable to understand these results in more detail analytically. 

The arrival of the sound pulse at the stellar surface is accompanied by at least three robust observational signatures:  (1)  The $\sim 0.2-1$ Mach number of the surface of the star leading up to the sound pulses arrival is likely accompanied by a modest decrease in the stellar effective temperature, though if the photosphere undergoes internal shocks in some WR models the change in emission is likely to be qualitatively different.   (2) Weak shock breakout emission.  (3)  Recombination powered emission associated with the unbound ejecta. \citet{lovegrove13}, \citet{piro13}, \& \citet{lovegrove17} quantify the second and third of these signatures for RSG progenitors while \citet{fernandez17} also quantify them for BSG and WR progenitors.    In addition to these robust signatures, the low energy shocks associated with failed supernovae are likely to lead to extended fallback accretion onto the black hole. This could power a variety of transients, particularly if there is sufficient angular momentum that an accretion disk forms (e.g., \citealt{Quataert_Kasen_2012,Woosley_Heger_2012,Dexter_Kasen_2013}).

The theoretical formalism developed in Section \ref{sec:general} is likely applicable to other problems of astrophysical interest. For example, the gravitational wave emission immediately following the inspiral of two supermassive black holes will result in a mass loss of a few percent of the sum of the initial masses. If there is a circumbinary disc present, the mass loss will induce a dynamical response of the disc, producing radial motions and shocks throughout the flow.  Indeed, this is observed in the simulations of \citet{rossi10} (though they find that the gravitational-wave induced kick of the remnant black hole can have a larger effect). Most work on this problem has focused on thin disks, but geometrically thick disks are likely present in the majority of galactic nuclei.  The spherical analysis of this paper -- or a modest extension that includes rotation -- may be particularly appropriate in these cases.

\section*{Acknowledgments}
We thank Stephen Ro for useful conversations.   ERC was supported by NASA through the Einstein Fellowship Program, grant PF6-170150. EQ was supported in part by a Simons Investigator award from the Simons Foundation, and the David and Lucile Packard Foundation. RF acknowledges support from NSERC of Canada and from the Faculty of Science at the University of Alberta. This work was also supported in part by the Gordon and Betty Moore Foundation through Grant GBMF5076.  We acknowledge stimulating workshops at Sky House and Oak Creek Ranch where these ideas germinated. This research used resources of the National Energy Research
Scientific Computing Center (NERSC), which is supported by the Office of
Science of the U.S. Department of Energy under Contract No. DE-AC02-05CH11231 (repository m2058).

\bibliographystyle{mnras}
\bibliography{references}

\appendix
\section{Numerical results for polytropes}
\label{sec:appendix}

\begin{figure}
\includegraphics[width=0.495\textwidth]{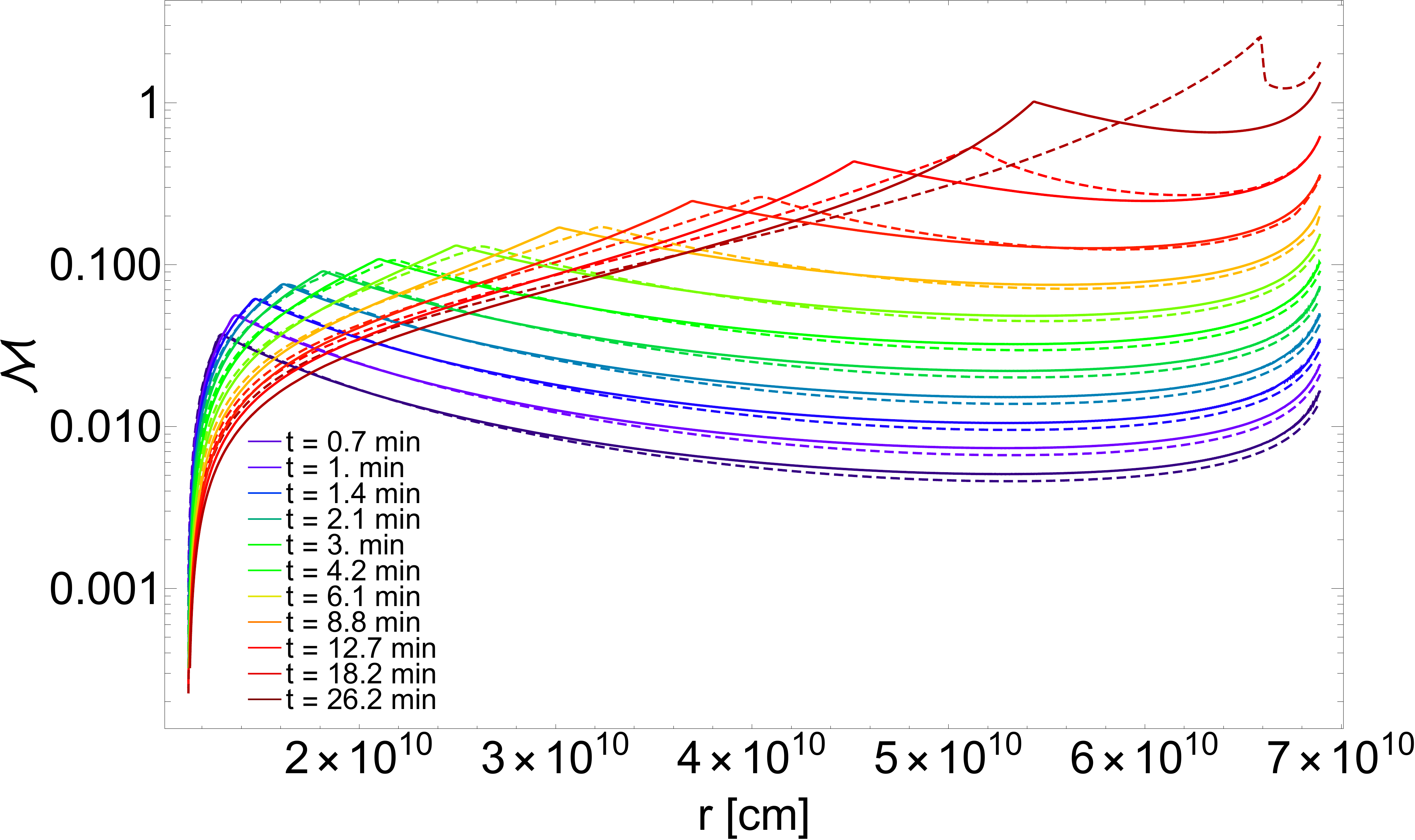}
\includegraphics[width=0.485\textwidth]{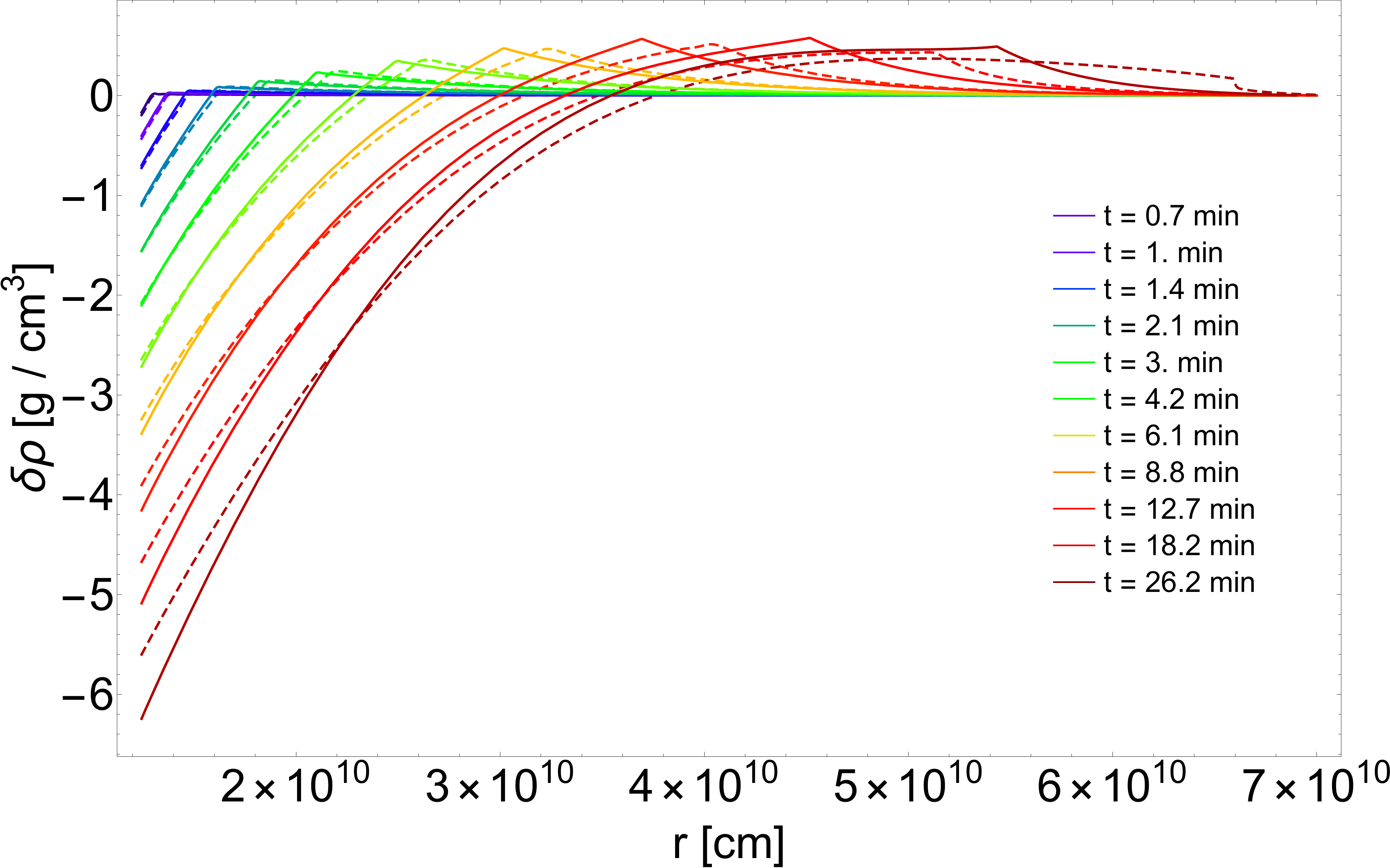}
\caption{The response of a $1M_{\odot}$, $1R_{\odot}$, $\gamma = 1.5$ polytrope when the mass interior to a dimensionless radius $\xi = 1$ is removed instantaneously. The top panel shows the Mach number, the bottom panel shows the difference in the time-dependent density and the initial, polytropic density profile, and the different curves correspond to the times in the legend. The solid curves show the analytic prediction, while the dashed curves give the results from a 1-D, Lagrangian hydrodynamics code.  The agreement is excellent so long as the linear assumption that $\mathcal{M} \lesssim 1$ is satisfied.}
\label{fig:numerics}
\end{figure}

To quantify the accuracy of the linear perturbation calculations used throughout the main text, we used a 1-D, Lagrangian hydrodynamics code to investigate numerically the response of a $1M_{\odot}$, $1R_{\odot}$, $\gamma = 1.5$ polytrope in which we instantaneously remove the mass interior to the dimensionless polytropic radius $\xi = 1$ (which corresponds to $\delta m \simeq 0.3$). 

The top panel of Figure \ref{fig:numerics} shows the Mach number of the flow induced by the mass loss, while the bottom panel shows the difference $\delta \rho$ between the initial, polytropic density profile and the time-dependent one, which is $\rho_1$ in the notation introduced in Section \ref{sec:general}. The different curves correspond to the times in the legend, with the solid curves being the analytic prediction following from Equation \eqref{fluxex} and the dashed curves the results from the numerical simulation. 

From this figure we see that the analytic approach established in Section \ref{sec:general} gives an excellent approximation to the full, nonlinear response of the envelope. Indeed, significant differences only arise once the Mach number of the flow approaches unity. The numerical solution shows that the pulse steepens and shocks (i.e., reaches a Mach number of 1) sooner and at a slightly larger radius in the envelope.

\bsp	
\label{lastpage}
\end{document}